\documentclass[journal]{IEEEtran}

\ifCLASSINFOpdf
   \usepackage[pdftex]{graphicx}
  
\else
  
\fi

\usepackage{mathtools, nccmath}
\usepackage{amsmath}
\usepackage{cite}
\usepackage{siunitx}
\usepackage{subcaption}
\usepackage{placeins}
\usepackage{booktabs}
\usepackage{enumitem}
\usepackage{eqnarray}
\usepackage{color}
\usepackage{amsfonts}
\usepackage{physics}
\newtheorem{thm}{Theorem}

\begin{document}

\title{Development of Robust Fractional-Order Reset Control}

\author{Linda~Chen,
        Niranjan~Saikumar,
        and~S.Hassan~HosseinNia, IEEE senior member% <-this % stops a space}
\thanks{Y.Y.L. Chen was with the Delft Center for Systems and Control, Delft University of Technology, Delft,
The Netherlands e-mail: lindachen93@gmail.com}% <-this % stops a space
\thanks{N. Saikumar and S.H. HosseinNia are with the Department of Precision and Microsystems Engineering, Delft University of Technology, Delft, The Netherlands. e-mail: n.saikumar@tudelft.nl, s.h.hosseinniakani@tudelft.nl}% <-this % stops a space

}

\markboth{Journal of \LaTeX\ Class Files,~Vol.~14, No.~8, December~2017}%
{Shell \MakeLowercase{\textit{et al.}}: Bare Demo of IEEEtran.cls for IEEE Journals}

\maketitle

\begin{abstract}
In this paper, a framework for the combination of robust fractional order CRONE control with non-linear reset is given for both first and second generation CRONE control. General design rules are derived and presented for these CRONE reset controllers. Within this framework, fractional order control allows for better tuning of the open-loop responses on the one hand. \color{black}{On the other, reset control enables a reduction in phase lag and a corresponding increase in phase margin compared to linear control for similar open loop gain profile. Hence, the combination of the two control methods can provide well-tuned open-loop responses that can overcome the fundamental linear control limitation of Bode's gain-phase relationship.} \color{black} Moreover, as established loop-shaping concepts are used in the controller design, CRONE reset can be highly compatible with the industry. The designed CRONE reset controllers are validated on a one degree-of-freedom Lorentz-actuated precision positioning stage. On this setup, CRONE reset control is shown to provide better tracking performance compared to linear CRONE control, which is in agreement with the predicted performance improvement.
\end{abstract}


\IEEEpeerreviewmaketitle

\section{Introduction}

\IEEEPARstart{M}{otion} control in (sub)nanometre precision positioning remains a present-day challenge in the high-tech industry. Advances in semiconductor manufacturing, production of micro- and nano-scale electronic devices (MEMS and NEMS) and imaging of nanostructures are among the applications which have increased demand for high degree of precision positioning systems. Conventional and popular PID controllers and even other linear controllers find it increasingly difficult to satisfy the demands in presence of uncertainties, which become more prominent when moving to smaller scales and higher bandwidths. In linear control, fundamental relations as the Bode's gain-phase relation and the waterbed effect \cite{skogestad2007multivariable} inevitably establish trade-offs between system performance in terms of reference-tracking, noise attenuation and disturbance rejection, and robustness. This makes precision positioning control an interesting problem from both system design and control point of view, considered for instance in \cite{schmidt2014design} and \cite{tan2007precision}.

In CRONE (Commande Robuste d'Ordre Non Entier - which translates to Non-Integer Order Robust Control) control {\cite{sabatier2015fractional} additional flexibility in the trade-off between robustness and performance is obtained using fractional operators. Although fundamental relations of linear control still apply, the fractional operators allow for better and easier tuning of required stability margins and open-loop shape. To overcome Bode's gain-phase relation and provide greater relief to the robustness-performance trade-off, non-linear reset control is considered.
	
Non-linear reset control has been the focus of many researchers in past and present years, starting from the first work of Clegg in 1958 \cite{clegg1958nonlinear}. In this work, a reset integrator (also known as Clegg Integrator (CI)) was introduced: an integrator which is reset when its error input equals zero. Using describing function analysis \cite{vidyasagar2002nonlinear}, it is seen that the reset integrator has a phase lag of only \SI{38}{\degree}, hence providing \SI{52}{\degree} phase lead with respect to a linear integrator for the same \SI{-20}{\deci\bel/}decade gain slope. Recognizing the potential of this profitable gain-phase characteristic, several works prove the improved performance using reset control, such as \cite{prieur2011} and \cite{witvoet2007}. Alternative resetting laws for improved robustness and/or performance have also been proposed. These include partial reset (non-zero after-reset state value), variable reset \cite{zheng2007}, both constant and variable reset band in \cite{banos2009stability} and \cite{VIDAL2010170} respectively, and (variable \cite{vidal2008pi}) reset percentages in the PI+CI compensator approach (PI controller with a reset integrator), for which a control design framework has been developed in \cite{banos2007definition}. Other resetting conditions include resetting at fixed time instants rather than fixed state values \cite{banos2011}, quadratic resetting conditions \cite{van2017frequency} and conditions obtained in an optimization problem \cite{Li2011optimal}. 

For these reset approaches, stability theorems have been developed. Generally, the works concerning stability proofs can be divided into Lyapunov-based and passivity based proofs \cite{Carrasco2010}. In the former, the $H_\beta$-condition \cite{banos2011reset} is one of the conditions with which one can prove stability for reset systems with stable linear base. In a recent work, sufficient stability conditions based on measured frequency responses are given \cite{van2017frequency}, which aims to eliminate the need for solving linear matrix inequalities (LMI) present in most of the previous stability conditions and thus making reset controllers more accessible to control engineers in industry. As a result of the existing research, many applications of reset control exists. Examples include applications in process control \cite{davo2013},\cite{perez2011}, positioning systems \cite{hazeleger2016},\cite{zheng2007} and hard-drive disks \cite{Li2011}, \cite{Li2009}. Still, reset control synthesis remains an actively researched topic.  

In the works of \cite{hosseinnia2013basic} and \cite{hosseinnia2014general} the framework for fractional order reset has already been mathematically founded. Several works in this field include generalization of the CI, PI+CI and FORE (first order reset element) to CI$^\alpha$ \cite{hosseinnia2014general}, PI$^\alpha$+CI$^\alpha$ \cite{hosseinnia2014general} and GFrORE \cite{niranjan2017} respectively. In \cite{hassan2015} a fractional order reset system with an iterative learning algorithm was proposed to increase robustness in the presence of model uncertainties and avoid limit cycles simultaneously. 

\color{black} {Although fractional order reset control exists in literature, reset applied to CRONE control specifically has not yet been done. The motivation for this work arises from the fact that a robust design methodology already exists with CRONE. However as noted earlier, CRONE controllers are linear and hence suffer from fundamental limitations. This paper aims to extend this design methodology to include reset actions to obtain further relief in the robustness-performance trade-off and provide new design rules for robust fractional order reset control. The preliminary work in this regard has been presented in \cite{LindaACC} with reset introduced into first generation CRONE controllers. However, second generation CRONE which can provide robust performance even in the case of non-asymptotic phase behaviour for the system in the region of bandwidth is of greater interest to the precision control community. The ideas for first generation CRONE are provided and then extended for second generation in this paper. Additionally, apart from the analysis of results obtained from second generation CRONE reset, the performance in terms of disturbance rejection for both generations has also been addressed in this paper.}

\color{black}

The paper is structured as follows. Section \ref{sec:preliminary} concerns fundamentals of CRONE control and reset control. Then follows the formulation of CRONE reset control and design rules in section \ref{sec:cronereset}. The practical application of designed CRONE reset controllers are given in section \ref{sec:practical}, followed by a discussion of experimental results in section \ref{sec:results}. Finally, conclusions are provided in section \ref{sec:conclusion}.

\section{Preliminaries}\label{sec:preliminary}
\subsection{Robust CRONE control}
The CRONE control framework provides a methodology for robust fractional order control design. Robustness is achieved by the creation of constant phase around open-loop bandwidth. This can be seen in Fig. \ref{img:bodecrone}. Under system gain deviations, robustness of the system is ensured as phase margin remains equal. Three generations of CRONE control have been formalized by \cite{sabatier2015fractional}. Only first two generations of CRONE are considered in this work: CRONE-1 and CRONE-2. CRONE-1 can be used for plants with asymptotic phase behaviour around the required bandwidth. CRONE-2 can be used for plants without this asymptotic phase behaviour. Both generations of CRONE provide robustness against gain deviation. CRONE-3 control uses complex fractional order, which is not practically implementable and thus not taken further into account.

\begin{figure*}[htbp]
\centering
\begin{subfigure}{0.49\linewidth}
\flushleft
\includegraphics[width=\linewidth]{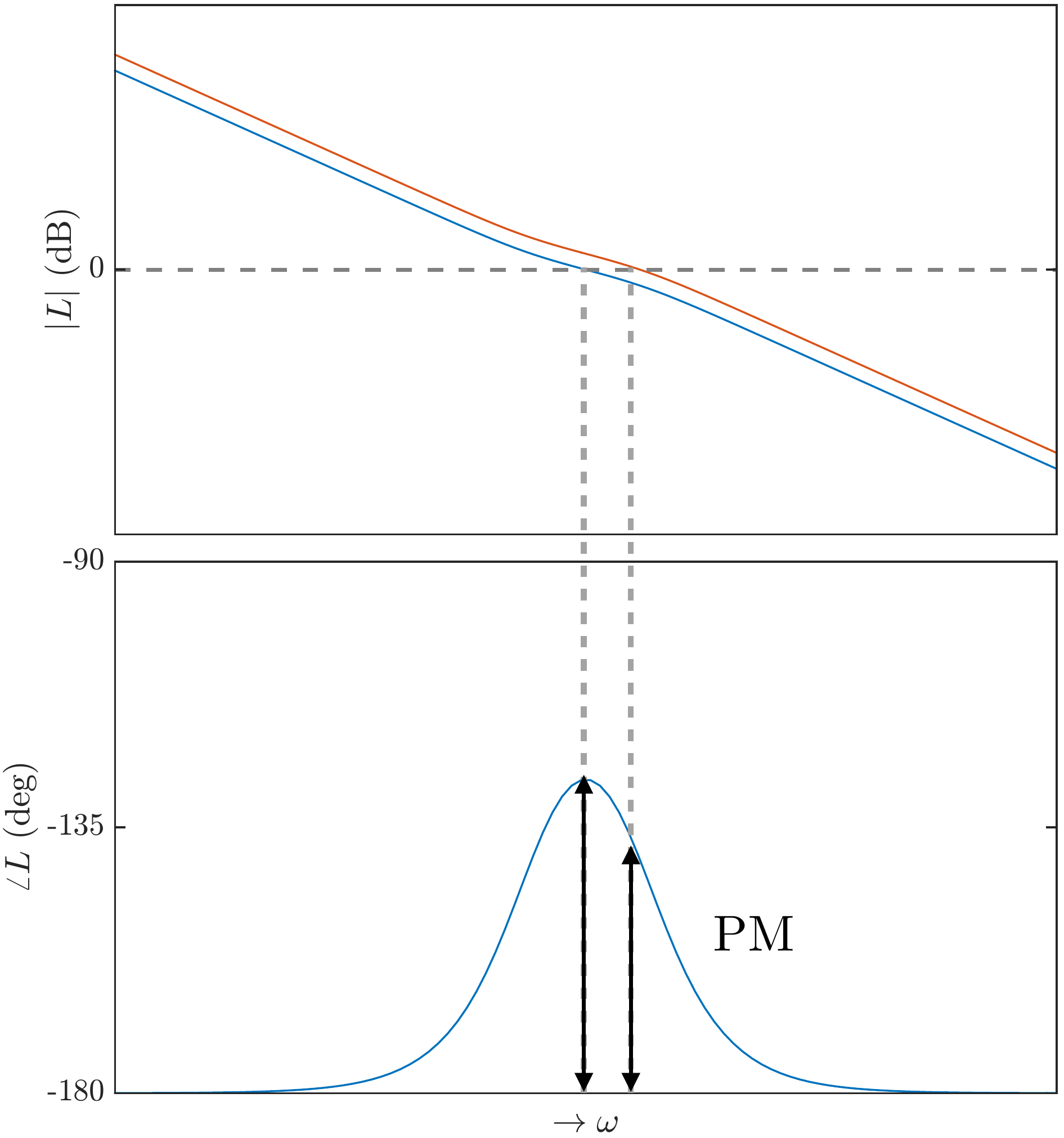}
\subcaption{}
\label{img:bodecrone1}
\end{subfigure}
\begin{subfigure}{0.49\linewidth}
\centering
\includegraphics[width=\linewidth]{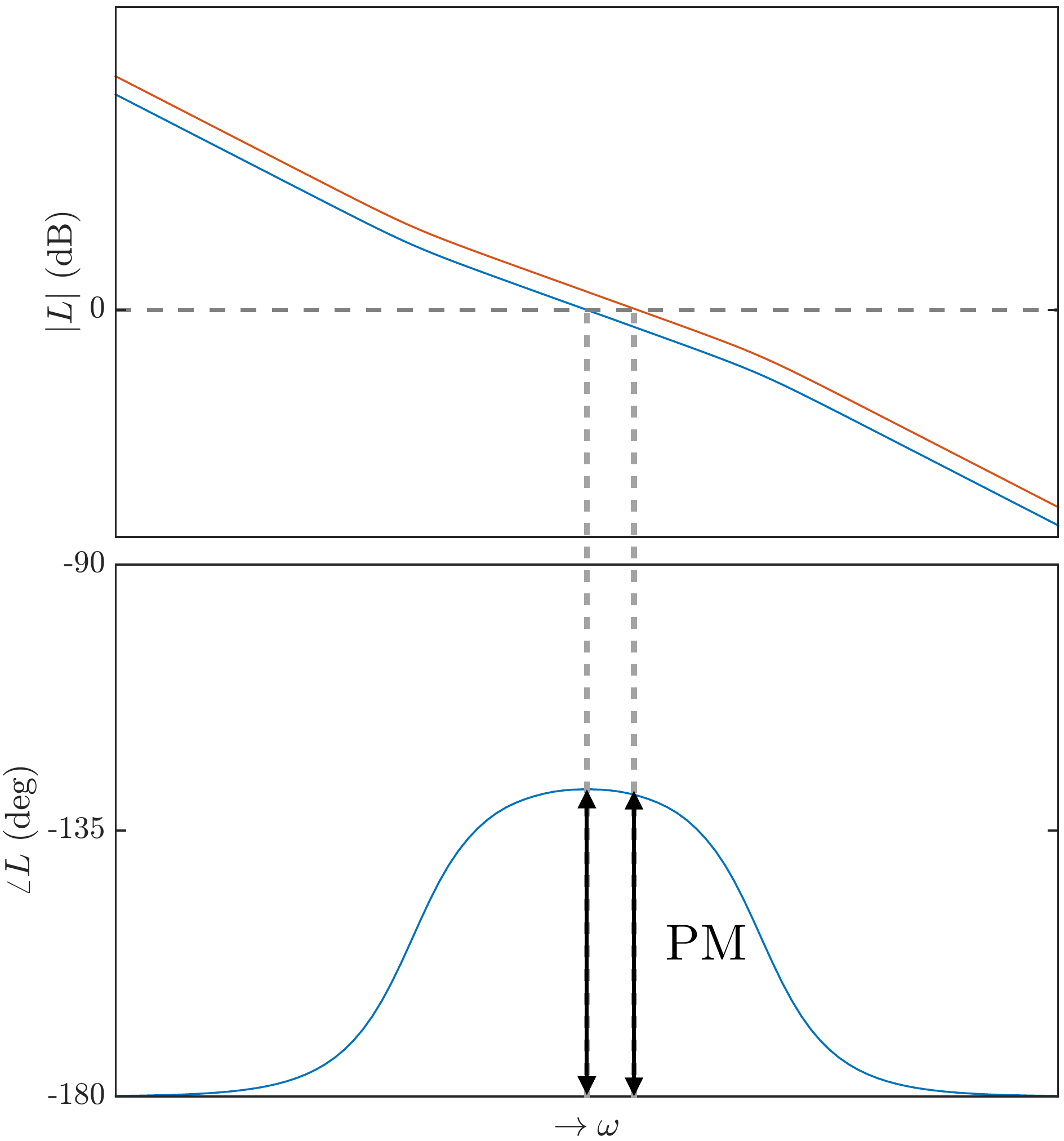}
\subcaption{}
\label{img:bodecrone2}
\end{subfigure}
\caption{(\subref{img:bodecrone1}) Typical open-loop response with conventional linear controllers. Under gain variations the phase margin (PM) fluctuates. (\subref{img:bodecrone2}) After achieving constant phase in the frequency range around bandwidth the phase margin is close to constant.}
\label{img:bodecrone}
\end{figure*}

\subsubsection{First generation CRONE}
A first generation CRONE controller, also referred to as CRONE-1, has a similar transfer function to an integer order series PID controller:
\begin{equation}\label{eq:Cf}
C_F(s)=C_0\bigg(1+\frac{\omega_I}{s}\bigg)^{n_I}\Bigg(\frac{1+\frac{s}{\omega_b}}{1+\frac{s}{\omega_h}}\Bigg)^\nu\frac{1}{\big(1+\frac{s}{\omega_F}\big)^{n_F}}
\end{equation}

with $\omega_I$ and $\omega_F$ being the integrator- and low pass filter corner frequencies, $\omega_b$ and $\omega_h$ the corner frequencies of the band-limited derivative action, $\nu \in \mathbb{R} \cap [0,1]$ the fractional order of the derivative action and $n_I,n_F \in \mathbb{N} $ being the order of the integrator and low pass filter respectively. The difference between a series integer order PID controller and a first generation CRONE controller is that the order $\nu$ is fractional instead of an integer, making first generation CRONE `a fractional PID controller'. The flat phase behaviour illustrated in Fig. \ref{img:bodecrone2} is created by choosing a wider frequency range in which the derivative action is active (compared to PID control) and by decreasing the order $\nu$ to a fractional value. 

The fractional order $\nu$ can be calculated from:

\begin{equation}\label{eq:alpha}
\nu=\\
\frac
{
	\begin{multlined}
	-\pi+M_\Phi-\arg G(j\omega_{cg})+n_F\arctan\frac{\omega_{cg}}{\omega_F}+\\n_I\Bigg(\frac{\pi}{2}-\arctan \frac{\omega_{cg}}{\omega_I}\Bigg)
	\end{multlined}
}
{\arctan\frac{\omega_{cg}}{\omega_b}-\arctan\frac{\omega_{cg}}{\omega_h}}
\end{equation}
where $G(j\omega)$ is the plant frequency response and $M_\Phi$ is the required phase margin. The gain $C_0$ is chosen such that the loop gain at frequency $\omega_{cg}$ is equal to 1. 

\subsubsection{Second generation CRONE}
In second generation CRONE, which is alternatively addressed as CRONE-2, the desired open-loop is firstly designed. The resulting controller is comprised of this desired open-loop in series with the plant inverse.

Desired open-loop $\beta_0(s)$ is given as:
\begin{equation}
\beta_0(s)=C_0\Bigg(1+\frac{\omega_I}{s}\Bigg)^{n_I}\Bigg(\frac{1+\frac{s}{\omega_b}}{1+\frac{s}{\omega_h}}\Bigg)^{-\nu}\frac{1}{\big(1+\frac{s}{\omega_F}\big)^{n_F}}
\end{equation}
where the order $\nu \in \mathbb{R} \cap [1,2] $ is again fractional and given by:
\begin{equation}
\nu=\frac{-\pi+M_\Phi+n_F\arctan\dfrac{\omega_{cg}}{\omega_F}+n_I\Bigg(\dfrac{\pi}{2}-\arctan\dfrac{\omega_{cg}}{\omega_I}\Bigg)}{\arctan\dfrac{\omega_{cg}}{\omega_h}-\arctan
\dfrac{\omega_{cg}}{\omega_b}}
\end{equation}

The second generation CRONE controller finally has the following structure:
\begin{equation}
C_\mathrm{S}(s)=G_0^{-1}(s)\beta_0(s)
\end{equation}
in which $G_0(s)$ is the nominal plant.

\color{black}
In both CRONE-1 and CRONE-2, the resulting fractional order derivative is approximated in the required frequency range using CRONE approximation \cite{sabatier2015fractional}. The resulting higher order integer order transfer function approximates the required fractional order derivative's frequency behaviour.
\color{black}

\subsection{Reset control}
A general reset system can be described using following impulsive differential equations, according to the formalism in \cite{banos2011reset}:
\begin{equation}\label{eq:IDESore}
\Sigma_R:=\begin{cases}
\dot{x}_R(t)=A_Rx_R(t)+B_Re(t)&\text{if } e(t)\neq 0,\\
%e(t)\neq0,\\
x_R(t^+)=A_\rho x_R(t)&\text{if } e(t)=0,\\
% e(t)=0\\
u(t)=C_Rx_R(t)+D_Re(t)& \\
\end{cases}
\end{equation}
where matrices $A_R, B_R, C_R, D_R$ are the base linear state-space matrices of the reset controller, $e(t)$ is the error between output and reference, $u(t)$ is the control input signal, $x_R(t)$ are the states with $x_R = [x_r^T \ \ x_{nr}^T]^T$ where $x_r$ are the $n_r$ number of states being reset and $x_{nr}$ are the $n_{nr}$ states not being reset with $n_r + n_{nr} = n_R$ (total number of states of feedback controller), and $A_\rho$ is the reset matrix. $A_\rho$ is designed as a diagonal matrix with the elements corresponding to states $x_{nr}$ equal to one.

When the frequency response of a reset system is approximated using describing function analysis, it is seen that phase lag is significantly reduced by the non-linearity.

\subsubsection{Describing function analysis}\label{sec:describingfunction}
The general describing function of a reset system as defined in \cite{guo2009frequency} is given by:
\begin{equation}
G_\mathrm{DF}(j\omega)=C_R(j\omega I-A_R)^{-1}B_R(I+j\Theta_D(\omega))+D_R
\end{equation}
where  $\Theta_D(\omega)$ is defined as:
\begin{equation}\label{eq:thetad}
\Theta_D(\omega)=-\frac{2\omega^2}{\pi}\Delta(\omega)[\Gamma_D(\omega)-\Lambda^{-1}(\omega)]
\end{equation}

The definitions of the set of equations used are given below:
\begin{eqnarray*}
\begin{cases}
\Lambda(\omega)=\omega^2I+A_R^2\\
\Delta(\omega)=I+e^{\frac{\pi}{\omega}A_R}\\
\Delta_D(\omega)=I+A_\rho e^{\frac{\pi}{\omega}A_R}\\
\Gamma_D(\omega)=\Delta^{-1}_D(\omega)A_\rho \Delta(\omega)\Lambda^{-1}(\omega)
\end{cases}
\end{eqnarray*}

\subsubsection{General stability analysis}\label{sec:stability}

\color{black}
The reset system given in (\ref{eq:IDESore}) can be represented in closed-loop as
\begin{equation}\label{eq:IDcl}
\begin{cases}
\dot{x}(t)=A_{cl}x(t) + B_{cl}w(t)&\text{if } x(t) \notin \mathcal{M}(t),\\
x(t^+)=A_{\rho_{cl}} x(t)&\text{if } x(t) \in \mathcal{M}(t),\\
% e(t)=0\\
u(t)=C_{cl}x(t) + d(t)\\
e(t) = w(t) - C_{cl}x(t)
\end{cases}
\end{equation}
where $x = [x^T_p \ \ x_R^T]^T$
\begin{eqnarray*}
A_{cl} = \begin{bmatrix}
A_p&B_pC_r\\-B_RC_p&A_R
\end{bmatrix},
B_{cl} = \begin{bmatrix}
	0\\B_R
\end{bmatrix}\\
A_{\rho_{cl}} = \begin{bmatrix}
	I_{n_p}&0\\0&A_\rho
\end{bmatrix},
C_{cl} = \begin{bmatrix}
	C_p&0
\end{bmatrix}
\end{eqnarray*}
with $A_p, B_p, C_p$ being the state space matrices of the plant to be controlled with $n_p$ number of states and reset surface $\mathcal{M}(t)$ is given as
\begin{equation*}
\mathcal{M}(t) = \begin{Bmatrix}
\xi \in \mathbb{R}^{n_p+n_r} : e(t) = 0, (I - A_R)\xi \neq 0
\end{Bmatrix}
\end{equation*}

\begin{thm}{\cite{beker2004fundamental}}\label{th:1}
Let $V: \mathbb{R}^n\rightarrow\mathbb{R}$ be a continuously differentiable, positive-definite, radially unbounded function such that
\begin{eqnarray}\label{eq:lyapunov}
\dot{V}(x):=\Big(\pdv{V}{x}\Big)^TA_{cl}x<0,&\text{if } x \neq 0,\\
\Delta V(x):=V(A_{\rho_{cl}} x)-V(x)\leq 0,& \text{if } x \in \mathcal{M} \label{eq:lyapunov2} 
\end{eqnarray}

Then the equilibrium point $x = 0$ is globally uniformly asymptotically stable.
\end{thm}

From this condition the authors of \cite{beker2004fundamental} obtained following theorem for proving quadratic stability:

\begin{thm}{\cite{beker2004fundamental}} \label{th:2}
The reset control system (\ref{eq:IDcl}) is said to satisfy the $H_\beta$ condition if there exists a constant $\beta\in\mathbb{R}^{n_r}$ and positive-definite $P_\rho\in\mathbb{R}^{n_r\times n_r}$, such that 
\begin{equation}
H_\beta(s) = \begin{bmatrix}
\beta C_p&O_{n_r\times n_{nr}}&P_\rho
\end{bmatrix}(sI - A_{cl})^{-1}
\begin{bmatrix}
O_{n_p\times n_r}\\O_{n_{nr}\times n_r}\\I_{n_r}
\end{bmatrix}
\end{equation}
is strictly positive real. The reset control system in (\ref{eq:IDESore}) is quadratically stable if and only if it satisfies the $H_\beta$-condition.
%the restricted Lyapunov equation
%\begin{eqnarray}\label{eq:lmi1}
%P>0,&A_{cl}^TP+PA_{cl}<0,\\
%&B_0^TP=C_0
%\end{eqnarray}
%has a solution for $P$, where $C_0$ and $B_0$ are defined by:
%\begin{eqnarray}
%C_0=\begin{bmatrix}
%\beta C_{p}&O_{n_r\times n_{nr}}&P_\rho
%\end{bmatrix},
%B_0=\begin{bmatrix}
%O_{n_{nrp}\times n_r}\\O_{n_{r_{nr}}\times n_r}\\I_{n_r}
%\end{bmatrix}\label{eq:lmi2}
%\end{eqnarray}
%and $n_{nr}$ is the number of non-reset states in $\Sigma^*_r$.
\end{thm}	

This $H_\beta$ condition has been used in this paper for stability analysis. However, this is not the only stability theorem for reset systems present in literature. For example, \cite{nevsic2008stability} provides conditions for $L_p$ stability for arbitrary $p \in [1, \infty)$. Since the focus of this paper is not on developing stability theorems, this is not discussed in greater detail.

\color{black}

\section{Robust CRONE reset control}\label{sec:cronereset}

CRONE control by itself is fundamentally limited as a linear controller. Thus being a robust controller, the system may under-perform in terms of tracking, disturbance rejection and noise attenuation as a result of fundamental trade-offs in linear control. It is in this scenario that non-linear reset can provide relief. The novel combination of CRONE and reset control, will be addressed as CRONE reset control. The CRONE reset control concept can be broken down into three steps:

\begin{enumerate}
\item Design of robust CRONE controller
\item Addition of phase around bandwidth with non-linear reset
\item Retuning of open-loop slope around bandwidth to improve open-loop shape for same phase margin
\end{enumerate}

\color{black}
The first step results in a linear controller which is robust and hence suffers in tracking and noise attenuation. The introduction of reset and the subsequent retuning results in robustness being retained with improvement in other performance characteristics. Above procedure is summarized in the open-loop responses depicted in Fig. \ref{fig:procedure}. In the final step as depicted in Fig. \ref{fig:cronestep3} it can be seen that open-loop gain has improved at both low and high frequencies (for better tracking and improved attenuation of noise respectively) with respect to the linear CRONE case in Fig. \ref{fig:cronestep1}. 
\color{black}

CRONE reset control design requires computation of a new slope around bandwidth $\nu^*$. This value differs, depending on the amount of phase added by resetting action, and thus varies for different reset strategies. In this section, firstly reset strategies are formulated, a CRONE reset control structure is established and the new design rules are given for calculation of slope $\nu^*$ for a selection of reset strategies for both CRONE-1 reset and CRONE-2 reset. Finally, the general stability analysis, as introduced in section \ref{sec:stability}, is adapted for the developed CRONE reset control framework.

\begin{figure*}[hbt]
	\centering
	\begin{subfigure}{0.32\linewidth}
		\includegraphics[width=\linewidth]{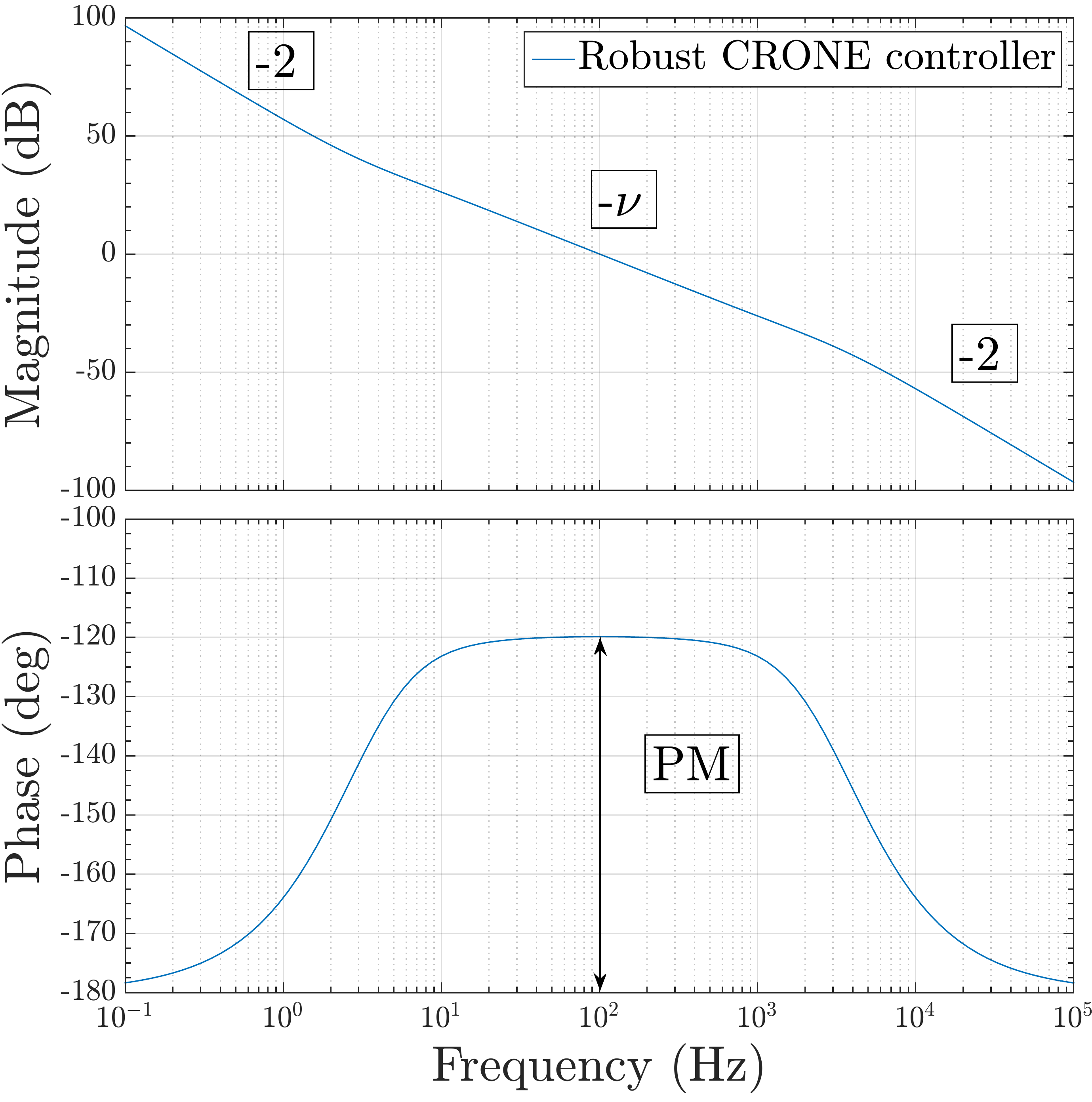}
		\caption{}
		\label{fig:cronestep1}
	\end{subfigure}
	\begin{subfigure}{0.32\linewidth}
		\includegraphics[width=\linewidth]{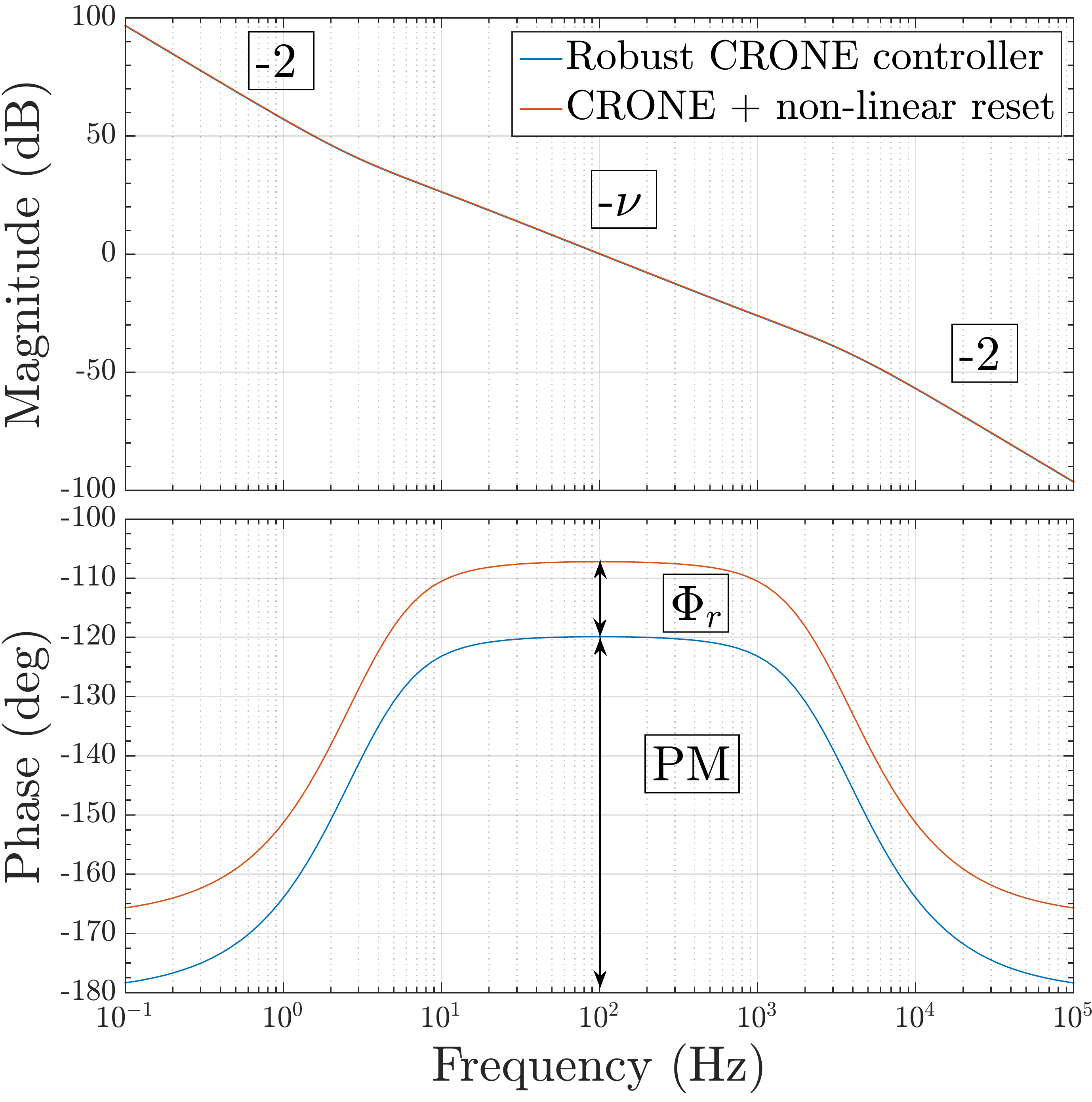}
		\caption{}
		\label{fig:cronestep2}
	\end{subfigure}
	\begin{subfigure}{0.32\linewidth}
		\includegraphics[width=\linewidth]{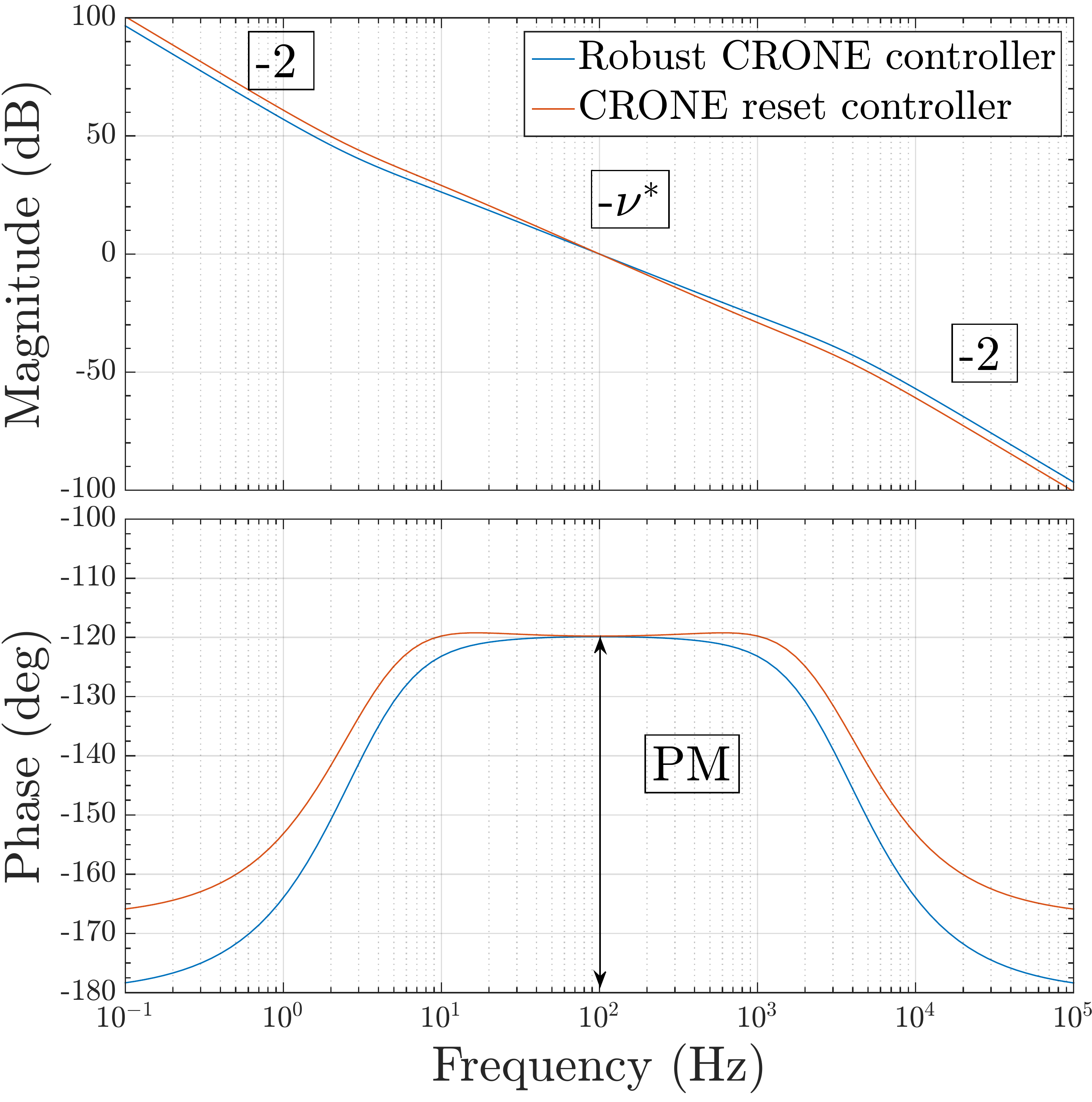}
		\caption{}
		\label{fig:cronestep3}
	\end{subfigure}
	\caption{Frequency domain open-loop responses showing CRONE reset control concept. Open-loop with indication of open-loop slopes for (\subref{fig:cronestep1}) step 1: robust CRONE controller, (\subref{fig:cronestep2}) step 2: additional non-linear reset adds phase $\Phi_r$, (\subref{fig:cronestep3}) step 3: improved open-loop shape with new fractional slope $\nu^*>\nu$. }
	\label{fig:procedure}
\end{figure*}

\subsection{Reset strategies}
We define reset strategy as a unique combination of choices for: base controller, part of controller transfer function to be reset, the order of the part to be reset and the selection of reset approaches. 
%Table \ref{tab:strategy} summarizes the available options for the first two generations of CRONE control and current reset approaches found in literature. 

%Integrator (wi/s), lead (...), and/or LPF (...) are the part of CRONE control (1) and (5) that can be chosen to reset while the rest will be kept as the linear part. These reset elements can have order 1 or higher. However in this paper we focus on first order reset elements. 
%As in litrature different approaches can be taken [cite] to reset i) Partial reset in which .... ii) reset percentage in which ...
%iii) variable reset in which ...
%iv) reset band where 
%However, in our paper the first strategies will be used.
%
%
%The low-pass filter reset part is disregarded since it adds phase significantly after the corner frequency, which does not fulfil the goal of adding phase around bandwidth. 

%To illustrate how a reset strategy is constructed, following transfer function shows the structure of $n$-th order CRONE-1 lead-lag reset.
Below the construction of a reset strategy is illustrated for CRONE-1 reset as an example:

\begin{equation}
C_{F}(s)=\Sigma_{r}\Sigma_{nr}
\end{equation}
in which $\Sigma_{nr}$ is the linear part and $\Sigma_r$ is the reset part of the transfer function as given in (\ref{eq:Cf}). The reset part could be one of the following:
\begin{enumerate}[label=\Roman*)]
\item  integrator part $\bigg(\dfrac{\omega_I}{s}\bigg)$
\item  lead/lag part $\bigg(\dfrac{1+\frac{s}{\omega_b}}{1+\frac{s}{\omega_h}}$, lead: $\omega_b<\omega_h$, lag: $\omega_b>\omega_h\bigg)$
\item  first order filter part $\Bigg(\dfrac{1}{1+\frac{s}{\omega_h}}$ or$\dfrac{1}{1+\frac{s}{\omega_l}}\Bigg)$
\end{enumerate} 
These reset elements can be of first order or higher. However in this paper we focus on first order reset elements only. 

\color{black}
With the second of 3 choices listed above, the lead and lag filters are assessed in two different ways. In the case of both lag and lead filters, resetting can be performed either only on the pole or on the pole-zero combination and results in unique frequency response. This is as shown in Table. \ref{tab:leadlag}. In the case where only the pole is reset as shown in the top row of Table. \ref{tab:leadlag}, the zero part is made proper by combining with the low pass filter part of the designed CRONE controller.
\color{black}

Comparing the amount of additional phase lag reduction seen, resetting of lag part is more favourable than resetting of lead part. Thus the remaining of the paper will focus on:
\begin{enumerate}[label=\Roman*)]
\item  integrator part ($\frac{\omega_i}{s}$)
\item  lag part 
\item  first order filter part ($\frac{1}{1+\frac{s}{\omega_b}}$)
\end{enumerate} 

\begin{table*}[htbp]
	\caption{Describing function for resetting different parts of a lead and a lag filter. Lag reset provides more phase at bandwidth than lead reset. Reset of first-order filter with lower corner frequency $\omega_b$ provides more phase at bandwidth than the same filter with higher corner frequency $\omega_l$.}
	\centering
	\begin{tabular}{p{2cm}cc}\\\toprule
		Reset part&lead filter $\Big(\frac{1+\frac{s}{\omega_l}}{1+\frac{s}{\omega_h}}\Big)$&lag filter $\Big(\frac{1+\frac{s}{\omega_h}}{1+\frac{s}{\omega_l}}\Big)$\\\midrule
		first order filter reset&\begin{minipage}{0.4\linewidth}
			\centering
			\includegraphics[width=\linewidth]{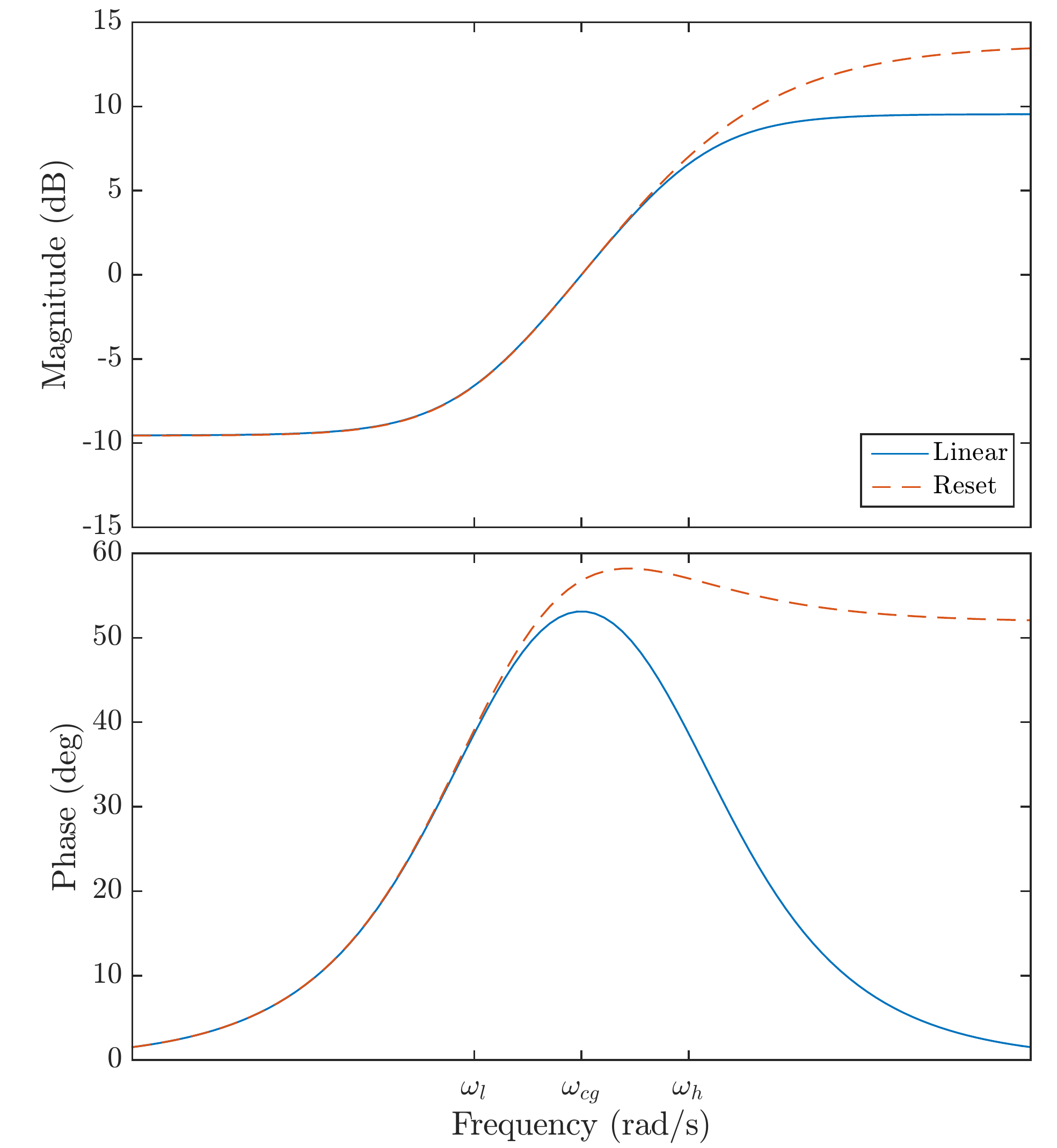}
			\\
			$\underbrace{(1+\frac{s}{\omega_l})}_{\Sigma_{nr}}\underbrace{\frac{1}{1+\frac{s}{\omega_h}}}_{\Sigma_r}$   
		\end{minipage}
		&\begin{minipage}{0.4\linewidth}
			\centering
			\includegraphics[width=\linewidth]{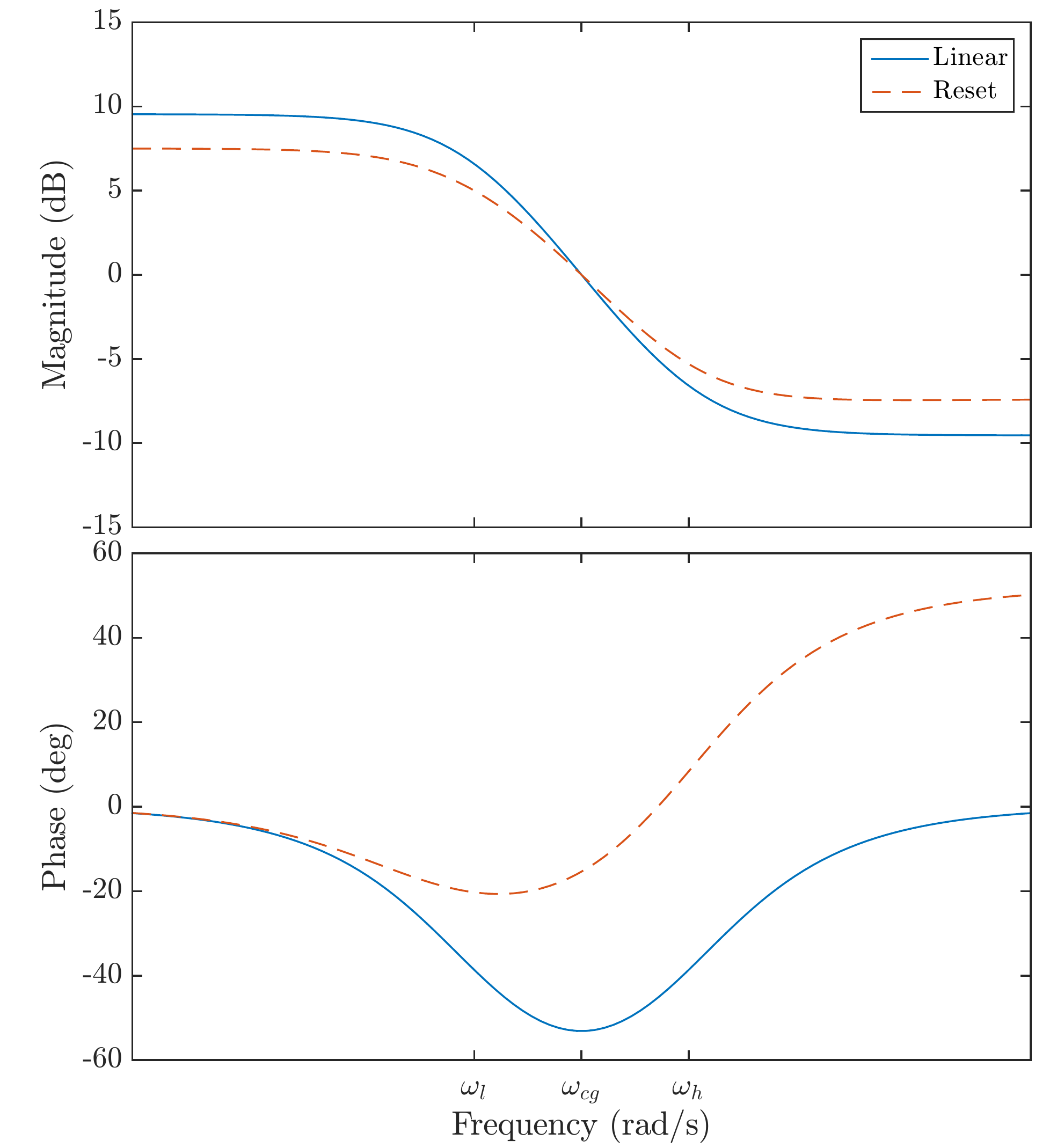}
			\\
			$\underbrace{(1+\frac{s}{\omega_h})}_{\Sigma_{nr}}\underbrace{\frac{1}{1+\frac{s}{\omega_l}}}_{\Sigma_r}$  
		\end{minipage}\\\hline
		lag/lead reset&\begin{minipage}{0.4\linewidth}
			\centering
			\includegraphics[width=\linewidth]{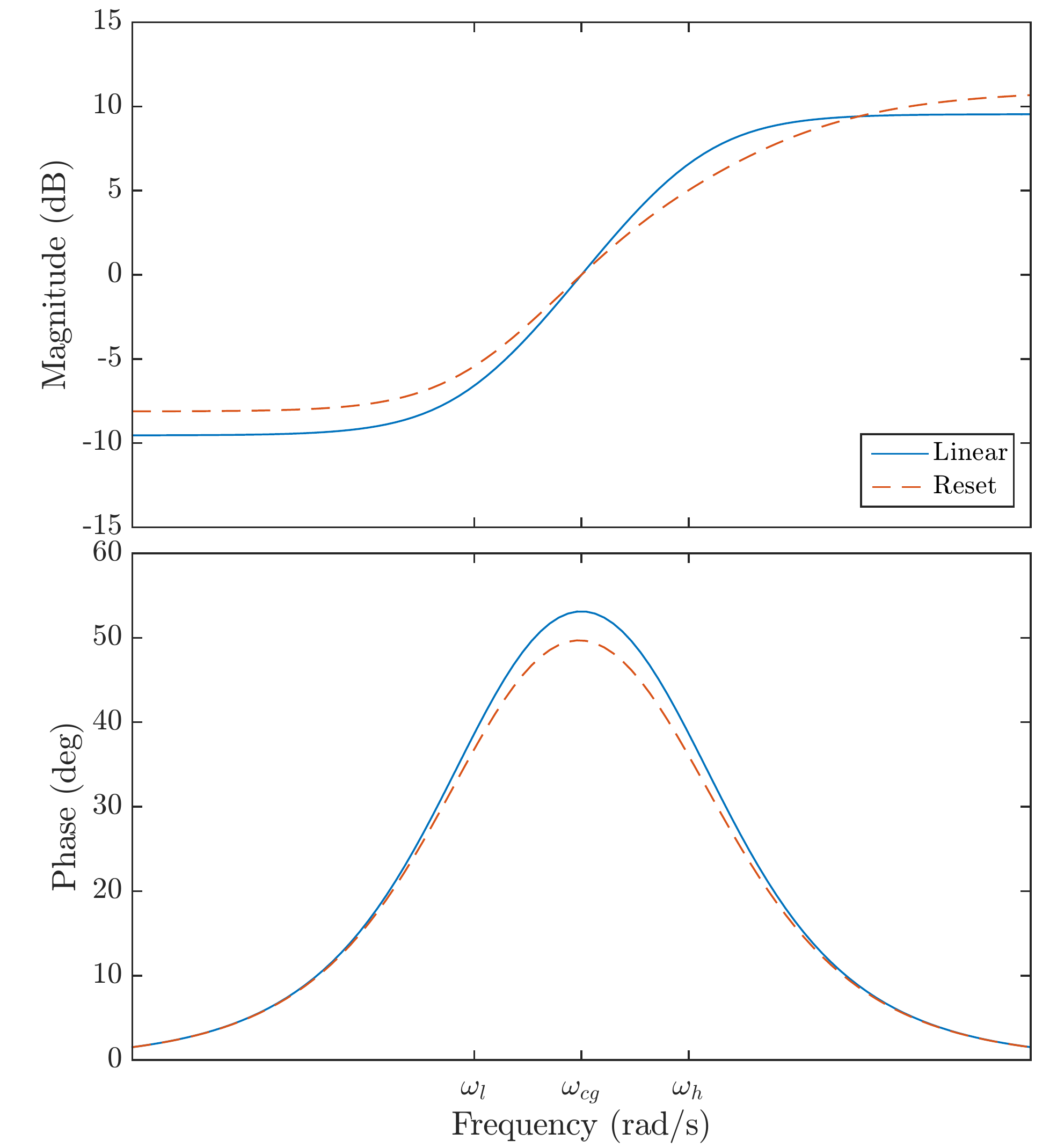}
			\\
			$\underbrace{\frac{1+\frac{s}{\omega_l}}{1+\frac{s}{\omega_h}}}_{\Sigma_r}$  
		\end{minipage}&\begin{minipage}{0.4\linewidth}
			\centering
			\includegraphics[width=\linewidth]{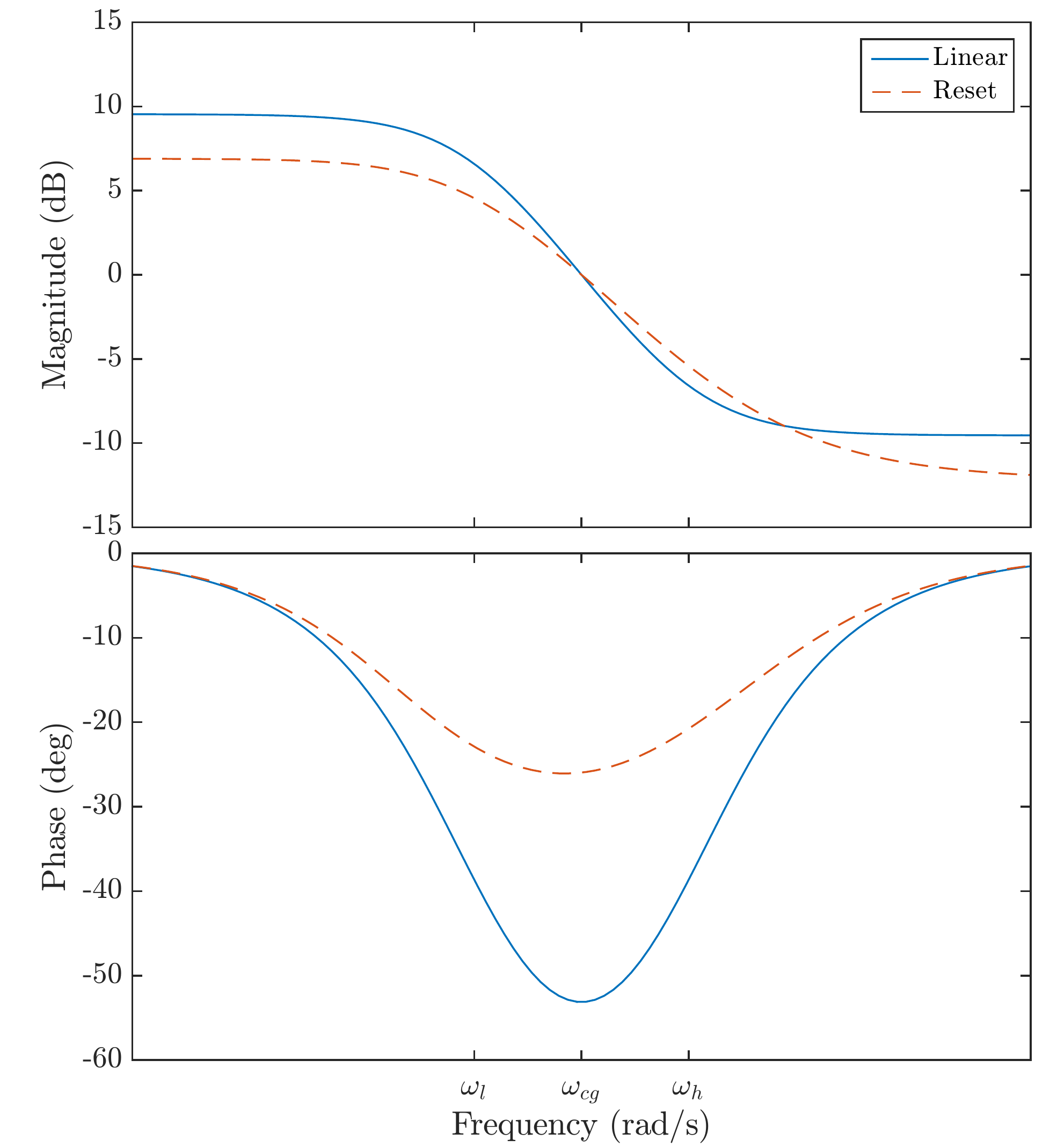}
			$\underbrace{\frac{1+\frac{s}{\omega_h}}{1+\frac{s}{\omega_l}}}_{\Sigma_r}$ 
		\end{minipage}\\\bottomrule
	\end{tabular}
	%\label{tab:DFleadlaglead}
	\label{tab:leadlag}
\end{table*}

As in literature, different approaches can be taken to reset \cite{banos2011reset}: 

\color{black}
\begin{enumerate}
\item partial reset -  where the state is not reset to zero resulting in the corresponding element of $A_\rho$ having a non-zero value.
\item reset percentage -  which uses the PI+CI compensator approach of having a PI loop in parallel with CI. Both loops have a weight assigned to them with the combined weight equalling one.
\item variable reset - where the non-zero value of partial reset and/or the weights of reset percentage are not fixed and can vary during operation.
\item reset band -  where reset is carried out when the error signal enters a band instead of zero-crossing.
\end{enumerate}

The describing function analysis in the case of variable reset and reset band is not straightforward and the equations provided in section \ref{sec:describingfunction} are not valid making design and analysis difficult. Also, these approaches are not robust and hence are not considered. Partial reset and reset percentage both provide ways to control the level of reset and hence nonlinearity and phase lag reduction achieved.
\color{black}

%
%
%\begin{table}[!h]
%\caption{CRONE reset strategies. Different CRONE reset strategies are possible by making combinations of choices per category/column. In the last column multiple choices are allowed.}
%\centering
%\begin{tabular}{p{2cm}p{2.7cm}p{1.8cm}p{2.5cm}}
%\toprule
%\bf{Base controller}&\bf{Reset part}&\bf{Reset order $n$}&\bf{Reset approaches}\\ \midrule
%\begin{minipage}{2\linewidth}
%\begin{enumerate}[leftmargin=*]
%\item CRONE-1
%\item CRONE-2
%\end{enumerate}
%\end{minipage}&
%\begin{minipage}{2\linewidth}
%\begin{enumerate}[leftmargin=*,label=\Roman*)]
%\item integrator $\frac{\omega_I}{s}$
%\item lead-lag $\frac{1+\frac{s}{\omega_b}}{1+\frac{s}{\omega_h}}$
%\item lag $\frac{1}{1+\frac{s}{\omega_h}}$
%%\item additional FORE
%%\item additional lead-lag
%\end{enumerate}\end{minipage}&
%\begin{minipage}{1.4\linewidth}
%\begin{enumerate}[leftmargin=*]
%\item first order
%\item second order
%\item higher order
%%\item fractional order
%\end{enumerate}
%\end{minipage}&
%\begin{minipage}{2\linewidth}
%\begin{itemize}[leftmargin=*]
%\item partial reset
%\item reset percentage
%\item variable reset
%\item reset band
%\end{itemize}
%\end{minipage}\\ \bottomrule
%
%\end{tabular}\label{tab:strategy}
%\end{table}

\subsection{Control structure}
The chosen reset approaches of partial reset and reset percentage constitute a CRONE reset controller with two degrees of freedom in tuning non-linearity in the system. The general state-space representation of the CRONE reset control system $\Sigma_R$ that consists of the reset part $\Sigma_r$ and non-reset part $\Sigma_{nr}$ is constituted as follows:

\begin{eqnarray}\label{eq:sysr}
&\Sigma_r:= & \begin{cases}
\dot{x}_r(t)=A_rx_r(t)+B_re(t)&\text{if } e(t)\neq0,\\
x_r(t^+)=\bar{A_\rho} x_r(t)&\text{if }  e(t)=0,\\
u_r(t)=C_rx_r(t)+D_re(t)& \\
\end{cases}\\
&\Sigma_{nr}:=& \begin{cases}
\dot{x}_{nr}(t)=A_{nr}x_{nr}(t)+B_{nr}u_r(t)\\
u_{nr}(t)=C_{nr}x_{nr}(t)+D_{nr}u_r(t) \\
\end{cases}\\
&\Sigma_R:= & \begin{cases}\label{eq:sysR}
\dot{x}_R(t)=A_Rx_R(t)+B_Re(t)&\text{if } e(t)\neq0,\\
x_R(t^+)=A_\rho x_R(t)&\text{if }  e(t)=0,\\
u(t)=C_Rx_R(t)+D_Re(t)& \\
\end{cases}
\end{eqnarray}
where $e(t)$ is the error signal, $u_r(t)$ is the output of $\Sigma_{r}$ which is in-turn input to the non-reset part, $x_r(t)$, $x_{nr}(t)$ and $x_R(t)=\begin{bmatrix}
x^T_r&x^T_{nr}\end{bmatrix}^T$ are the reset-controller states, non-reset controller states and CRONE reset controller states respectively and $A_\rho$, $\bar{A}_\rho$ are the reset matrices. Matrices $A_R, B_R, C_R, D_R$ are the base linear state-space matrices of the reset system, defined as:
\begin{eqnarray}
&A_R=&\begin{bmatrix}\label{eq:ar}
A_r&O\\
B_{nr}C_r&A_{nr}
\end{bmatrix},\\
&B_R=&\begin{bmatrix}
B_r\\B_{nr}D_r
\end{bmatrix},\\
&C_R=&\begin{bmatrix}\label{eq:cr}
D_{nr}C_r&C_{nr}
\end{bmatrix},\\
&D_R=&D_{nr}D_r
\end{eqnarray}

and reset matrix $A_\rho$ is defined as:
\begin{equation}\label{eq:bararho}
A_\rho=\mathrm{diag}(
\bar{A_\rho},I_{n_{nr}})
\end{equation}

The structure of this controller $\Sigma_R$ is shown in Fig. \ref{fig:rcontrol}. 

\begin{figure}[htbp]
\centering
\includegraphics[width=\linewidth]{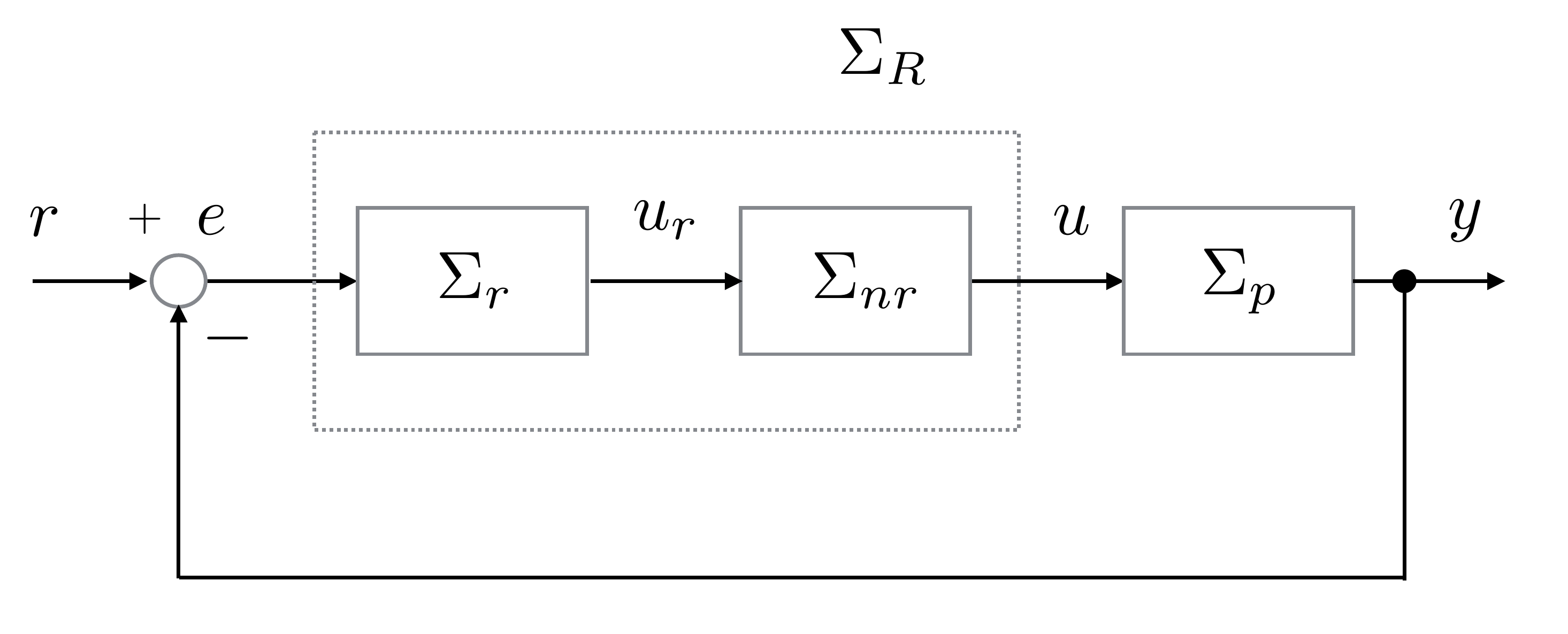}
\caption{Control structure of the CRONE reset controller $\Sigma_R$, which contains a linear part $\Sigma_{nr}$ and a non-linear reset part $\Sigma_r$. $\Sigma_p$ is the plant being controlled.}
\label{fig:rcontrol}
\end{figure}

%\subsection{Two-degree-of-freedom non-linearity tunable structure}
Slight alterations are made to the system definitions from (\ref{eq:sysr}) to (\ref{eq:sysR}) to obtain two degrees of freedom in tuning non-linearity within the system. The first chosen reset approach of partial reset establishes the first degree of freedom; reset matrix $\bar{A_\rho}$ is taken as:
\begin{equation}
\bar{A_\rho}=\gamma I_{n_r}
\end{equation} 
where $I_{n_r}$ is an identity matrix of size $n_r$ with $n_r$ being the number of reset states.  With this reset matrix the after-reset state value $x_r(t^+)$ is a fraction $\gamma$ of the before-reset state value $x_r(t)$. When $\gamma=0$ traditional reset occurs, whereas the system simplifies to a full linear system when $\gamma=1$. The conditions for open-loop stability of such a reset controller with non-zero resetting matrix is provided in \cite{guo2009frequency}.

\begin{figure}[htbp]
\centering
\includegraphics[width=\linewidth]{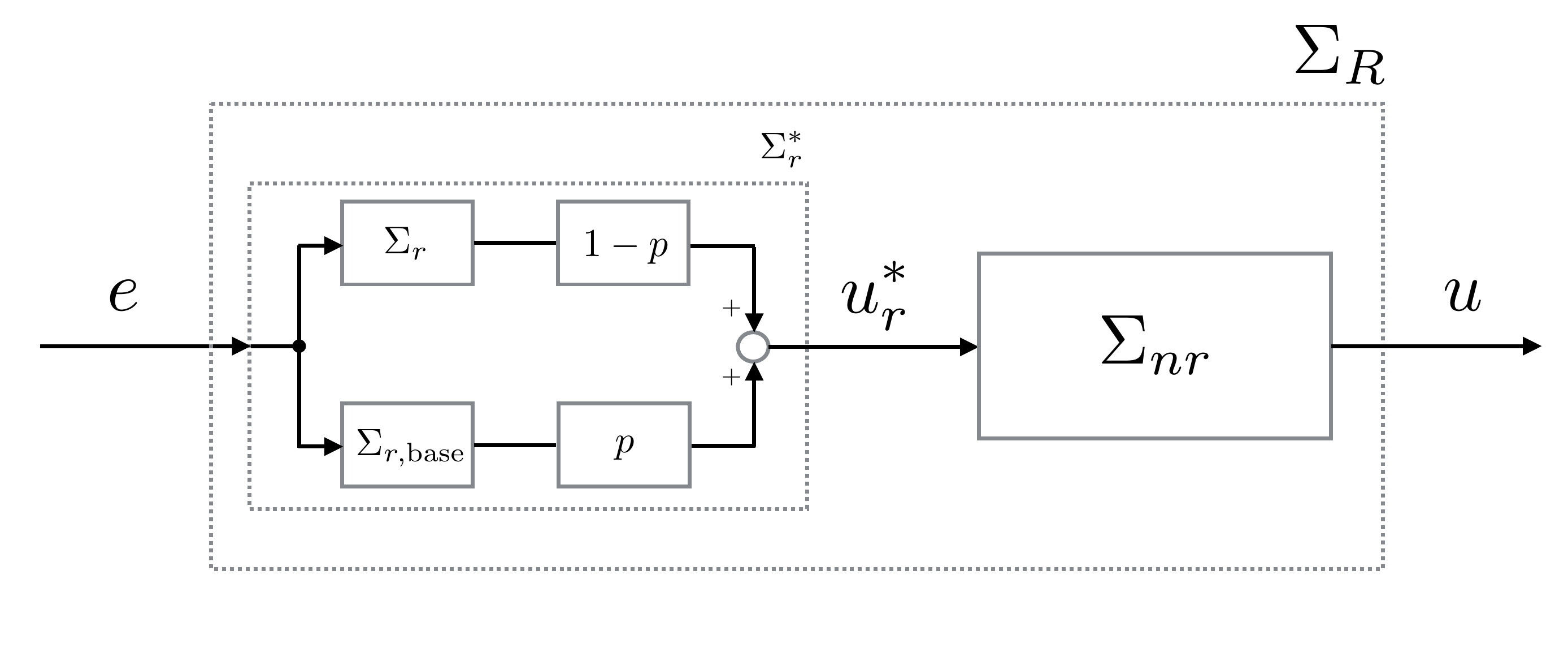}
\caption{Structure of the CRONE reset controller $\Sigma_R$ with two-degree-of-freedom non-linearity tuning, which contains a linear part $\Sigma_{nr}$ and a non-linear reset part $\Sigma_r$ with $A_\rho=\gamma I$. $\Sigma_{r,\mathrm{base}}$ is the base linear system of $\Sigma_r$.}
\label{fig:croneresetblock}
\end{figure}

The second chosen reset approach of reset percentage forms the second degree of freedom in tuning non-linearity in the system. A convex combination between the reset part $\Sigma_r$ and its linear base system $\Sigma_{r,\mathrm{base}}$ is taken as shown in figure \ref{fig:croneresetblock} in $\Sigma^*_r$. $p$ is the percentage of linearity in the system. When $p=0$, $\Sigma^*_r$ is equivalent to $\Sigma_r$ and when $p=1$ the system becomes fully linear.

The describing function of both reset approaches applied to a reset integrator is shown in Fig. \ref{fig:DFtuning}. It is evident that with both reset approaches the non-linearity in the system can be tuned and the amount of reset phase lead can be adjusted.

\begin{figure*}[htbp]
\centering
\begin{subfigure}{0.45\linewidth}
\includegraphics[width=\linewidth]{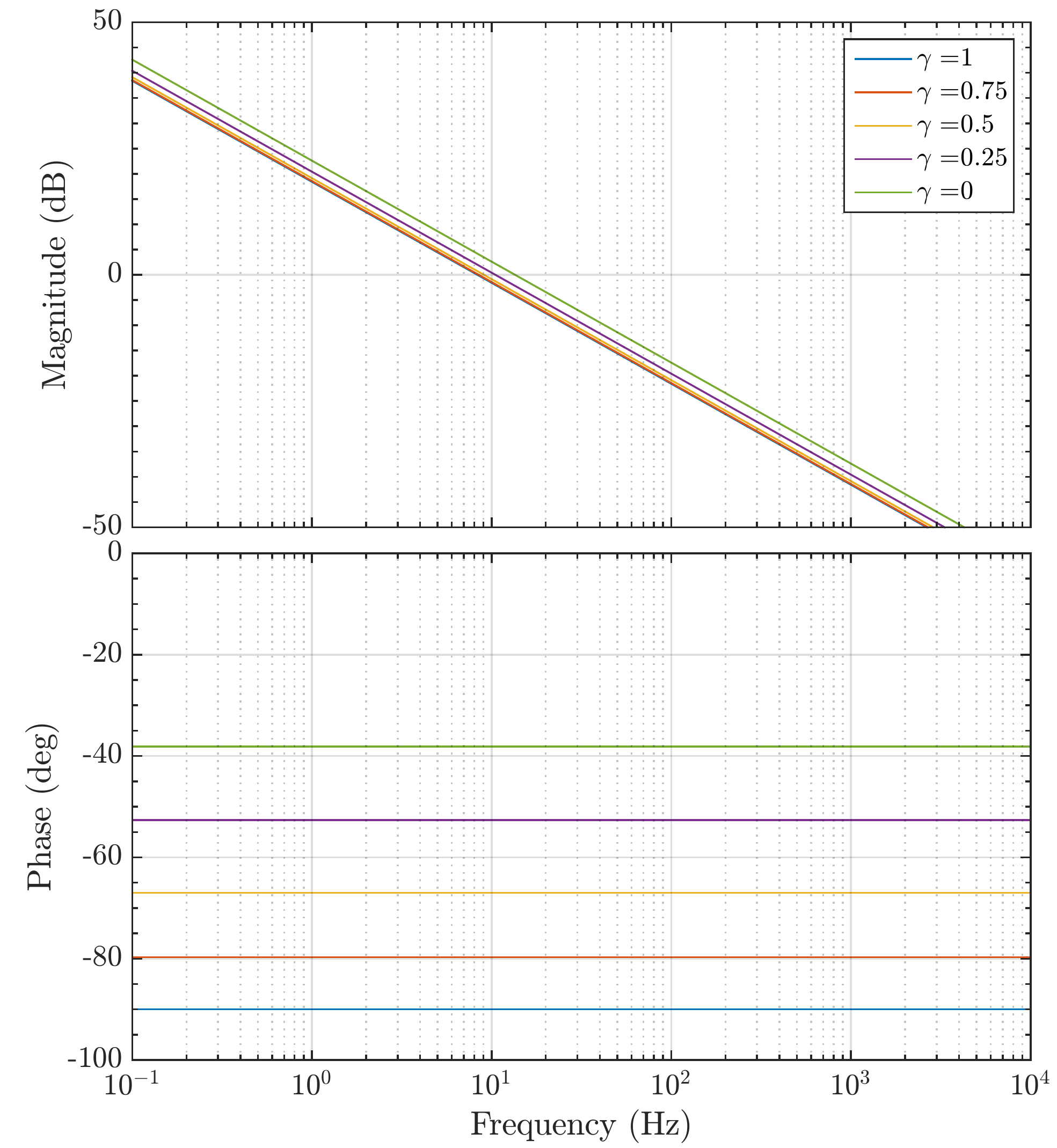}
\caption{}
\label{fig:DFgamma}
\end{subfigure}
\begin{subfigure}{0.45\linewidth}
\includegraphics[width=\linewidth]{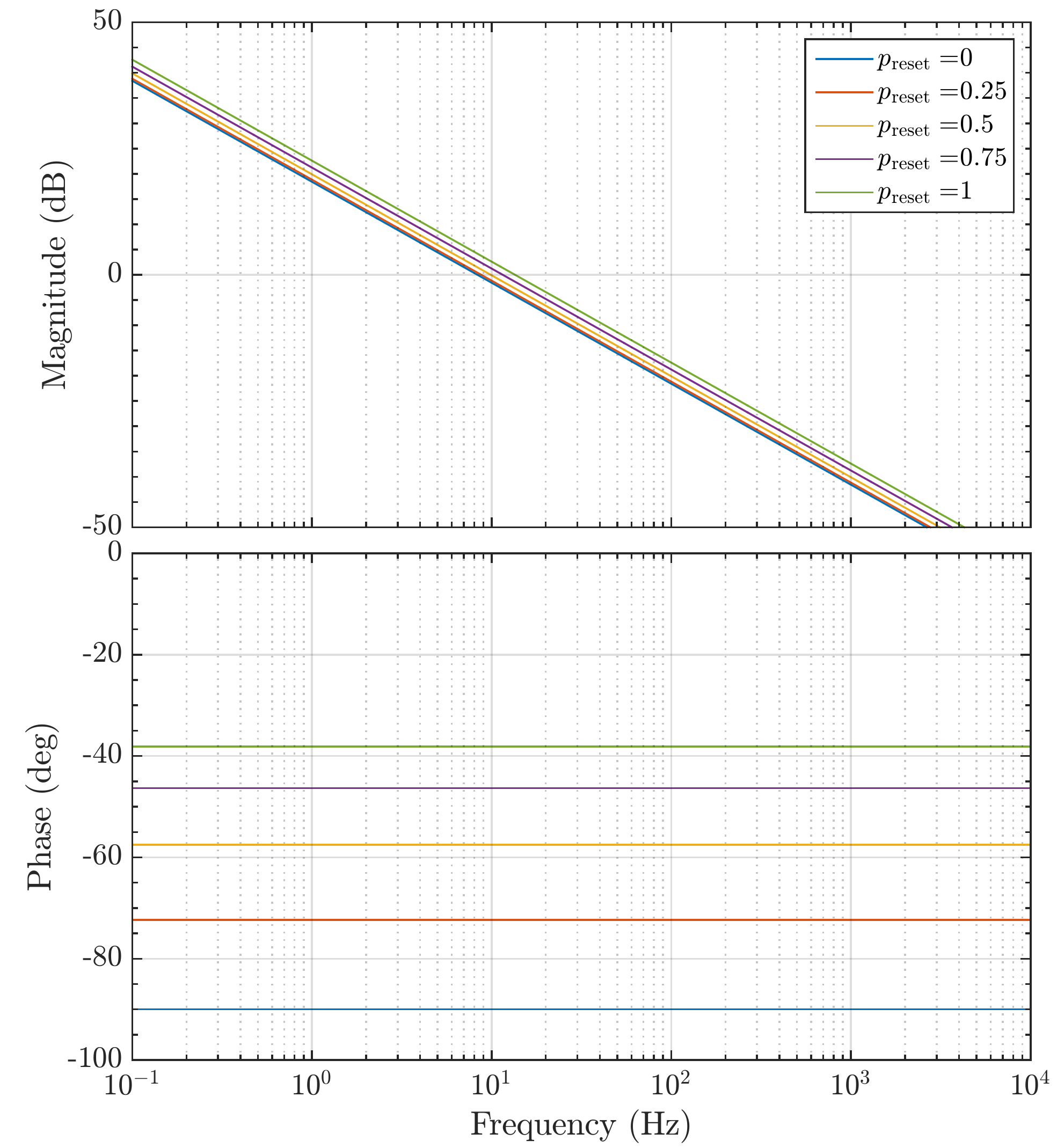}
\caption{}
\label{fig:DFpreset}
\end{subfigure}
\caption{Describing function of a reset integrator when tuning non-linearity (\subref{fig:DFgamma}) for different values of $\gamma$ with $p = 0$ and (\subref{fig:DFpreset}) for different values of $p$ with $\gamma = 0$.}
\label{fig:DFtuning}
\end{figure*}

\subsection{New design rules}

\color{black}
As noted at the beginning of this section, the design of CRONE reset controller consists of three steps with the first step being the design of linear CRONE controller. This is followed by a selection of reset strategy involving choice over $\Sigma_{r}$ and choice of control structure variables $\gamma$ and $p$. With these choices, the application of reset results in reduction in phase lag compared to its linear counterpart. This reduction which can be seen as phase lead achieved through reset can be calculated using describing function analysis for both CRONE-1 reset and CRONE-2 reset control.
\color{black}

%Firstly it is shown below how the reset phase lead is calculated from the describing function and linear transfer function. Then the reset phase lead is given for different reset strategies within CRONE-1 reset and CRONE-2 reset control. 
\subsubsection{Derivation reset phase lead}
%%In this method transient response is neglected. 
a linear combination is made between describing function of the reset system and its linear base equivalent to include the convex-combination structure with reset percentage $p$:
\begin{equation}\label{eq:DFanalytical}
\begin{split}
G^*_\mathrm{DF}(j\omega)= &p\Big(C(j\omega I-A)^{-1}B+D\Big)+\\
& (1-p)\Big(C(j\omega I-A)^{-1}B(I+j\Theta_D(\omega))+D\Big)
\end{split}
\end{equation}

The additional phase at bandwidth is given by $\Phi_r(\omega_{cg})$, which is retrieved by filling in $\omega=\omega_{cg}$ in:
\begin{equation}\label{eq:phir}
\Phi_r(\omega)=\angle G^*_\mathrm{DF}(j\omega)-\angle G(j\omega)
\end{equation}

\subsubsection{Reset strategy-dependent phase lead}
The amount of phase lead achieved by resetting is different depending on the chosen reset strategy. The reset phase leads are directly derived from (\ref{eq:DFanalytical}) and (\ref{eq:phir}) and given below for different reset strategies.

\paragraph{CRONE integrator reset}
 The reset phase lead for an integrator reset is:
\begin{equation}\label{eq:croneintphase}
\Phi_{r,\mathrm{int}}(\gamma,p)=\arctan \bigg(\frac{4}{\pi}(1-p)\frac{1-\gamma}{1+\gamma}\bigg)
\end{equation}
which is frequency-independent.

\paragraph{CRONE lead/lag reset}
As a base filter to define lead/lag reset strategy and finally acquire the controller representation in the form of  (\ref{eq:sysr}}) to (\ref{eq:sysR}), first consider following filter:
\begin{equation}\label{eq:Hllag}
H_\mathrm{ll}(s)=\frac{1+\frac{s}{a}}{1+\frac{s}{b}}
\end{equation}
where $a$ and $b$ are the corner frequencies. For such first order filters $\Theta_D(\omega)$, as in (\ref{eq:thetad}), is expressed as:
\begin{equation}
\Theta_D(\omega,b,\gamma)=\frac{2}{\pi}\frac{1+e^{-\pi\frac{b}{\omega}}}{1+(\frac{b}{\omega})^2}\frac{1-\gamma}{1+\gamma e^{-\pi\frac{b}{\omega}}}
\end{equation}
Then the phase lead for $H_\mathrm{ll}(s)$ in (\ref{eq:Hllag})
with partial reset $\gamma$ and reset percentage $p$ can be computed using (\ref{eq:phir}) as:
\begin{equation}\label{eq:phirllag}
\begin{aligned}
& \Phi_{r,\mathrm{ll}	}(\omega,a,b,\gamma,p)\\
& =\arctan\Bigg(\frac{(1-p)\Theta_D(\omega,b,\gamma)(1-\frac{b}{a})}{1+(\frac{\omega}{a})^2+(1-p)\frac{\omega}{a}\Theta_D(\omega,b,\gamma)(1-\frac{b}{a})}\Bigg)
\end{aligned}
\end{equation}

%For CRONE-1 $a=\omega_b$ and $b=\omega_h$ and for CRONE-2 it is the other way around. 
%The derivation of both $\Theta_D(\omega)$ and this phase lead can be found in \autoref{app:phase}.

\paragraph{CRONE first order filter reset}
When in (\ref{eq:Hllag})  $a\rightarrow \infty$  $H_{ll}(s)$ simplifies to a first order filter. Therefore the phase lead of (\ref{eq:phirllag}) becomes:
\begin{equation}\label{eq:cronelagphase}
\Phi_{r,\mathrm{fof}}(\omega,b,\gamma,p)=\arctan\bigg((1-p)\Theta_D(\omega,b,\gamma)\bigg)
\end{equation}

%For CRONE-1 $b=\omega_h$ and for CRONE-2 $b=\omega_b$.

Using the phase leads achieved with resetting in (\ref{eq:croneintphase})-(\ref{eq:cronelagphase}) for different reset strategies and applying (\ref{eq:nucrone}) and (\ref{eq:nucrone2}), the reset strategies applied to CRONE-1 and CRONE-2 are illustrated in the following subsections.	

\subsubsection{CRONE-1 reset}
Applying the new design rules to specific reset strategies for CRONE-1 results in following:
\paragraph{CRONE-1 integrator reset}
\begin{equation}
\begin{aligned}
& C_\mathrm{int}(s)\\
& =\underbrace{(\frac{\omega_I}{s} )^{n_I-1}C_0(\frac{s}{\omega_I}+1)^{n_I}\bigg(\frac{1+\frac{s}{\omega_b}}{1+\frac{s}{\omega_h}}\bigg)^\nu\frac{1}{(1+\frac{s}{\omega_F})^{n_F}}}_{\Sigma_{nr}}\underbrace{\frac{\omega_I}{s}}_{\Sigma_r}
\end{aligned}
\end{equation}
and phase lead with integrator reset is given by (\ref{eq:croneintphase}).
\paragraph{CRONE-1 lag reset}
\begin{equation}
C_\mathrm{lag}(s)=\underbrace{\bigg(\frac{1+\frac{s}{\omega_b}}{1+\frac{s}{\omega_h}}\bigg)^{\nu+1} C_0(1+\frac{\omega_I}{s})^{n_I}\frac{1}{(1+\frac{s}{\omega_F})^{n_F}}}_{\Sigma_{nr}}\underbrace{\frac{1+\frac{s}{\omega_h}}{1+\frac{s}{\omega_b}}}_{\Sigma_r}
\end{equation}
and phase lead with lag reset is given by (\ref{eq:phirllag}) with $b=\omega_b$ and $a=\omega_h$.

In the case of CRONE-1, it should be noted that the linear CRONE-1 has a lead element since the structure of CRONE-1 is similar to that of PID. However, according to Table. \ref{tab:leadlag}, since resetting lag part results in more lag reduction, a lag element is introduced with a corresponding

\paragraph{CRONE-1 first order filter reset}
\begin{equation}
C_\mathrm{fof}(s)=\underbrace{\frac{(1+\frac{s}{\omega_b})^{\nu+1}}{(1+\frac{s}{\omega_h})^\nu }C_0(1+\frac{\omega_I}{s})^{n_I}\frac{1}{(1+\frac{s}{\omega_F})^{n_F}}}_{\Sigma_{nr}}\underbrace{\frac{1}{1+\frac{s}{\omega_b}}}_{\Sigma_r}
\end{equation}
and phase lead with first order filter reset is given by (\ref{eq:cronelagphase}) with $b=\omega_b$.
\subsubsection{CRONE-2 reset}
Results for CRONE-2 reset are retrieved as:

\paragraph{CRONE-2 integrator reset}
\begin{equation}
\begin{aligned}
&\beta_{0,\mathrm{int}}(s) \\
& =\underbrace{(\frac{\omega_I}{s} )^{n_I-1}C_0(\frac{s}{\omega_I}+1)^{n_I}\bigg(\frac{1+\frac{s}{\omega_b}}{1+\frac{s}{\omega_h}}\bigg)^{-\nu}\frac{1}{(1+\frac{s}{\omega_F})^{n_F}}}_{\Sigma_{nr}}\underbrace{\frac{\omega_I}{s}}_{\Sigma_r}
\end{aligned}
\end{equation}
and phase lead with integrator reset is given by (\ref{eq:croneintphase}).
\paragraph{CRONE-2 lag reset}
\begin{equation}
\begin{aligned}
& \beta_{0,\mathrm{lag}}(s)\\
& =\underbrace{\bigg(\frac{1+\frac{s}{\omega_b}}{1+\frac{s}{\omega_h}}\bigg)^{-(\nu-1)} C_0(1+\frac{\omega_I}{s})^{n_I}\frac{1}{(1+\frac{s}{\omega_F})^{n_F}}}_{\Sigma_{nr}}\underbrace{\frac{1+\frac{s}{\omega_h}}{1+\frac{s}{\omega_b}}}_{\Sigma_r}
\end{aligned}
\end{equation}
and phase lead with lag reset is given by (\ref{eq:phirllag}) with $b=\omega_b$ and $a=\omega_h$.

\paragraph{CRONE-2 first order filter reset}
\begin{equation}
\begin{aligned}
& \beta_{0,\mathrm{fof}}(s)\\
& =\underbrace{\frac{(1+\frac{s}{\omega_b})^{-(\nu-1)}}{(1+\frac{s}{\omega_h})^{-\nu} }C_0(1+\frac{\omega_I}{s})^{n_I}\frac{1}{(1+\frac{s}{\omega_F})^{n_F}}}_{\Sigma_{nr}}\underbrace{\frac{1}{1+\frac{s}{\omega_b}}}_{\Sigma_r}
\end{aligned}
\end{equation}
and phase lead with first order filter reset is given by (\ref{eq:cronelagphase}) with $b=\omega_b$.

Once the phase lead achieved through application of reset is calculated for the chosen reset strategy and reset variables, new slope $\nu^*$ as a function of reset phase lead at bandwidth frequency $\Phi_r(\omega_{cg})$ can be calculated as:
\begin{equation}\label{eq:nucrone}
\nu^*=
\frac
{
	\begin{multlined}
	-\pi+M_\Phi-\angle G(j\omega_{cg})+n_F\arctan\frac{\omega_{cg}}{\omega_F}+\\n_I(\frac{\pi}{2}-\arctan\frac{\omega_{cg}}{\omega_I})-\Phi_r(\omega_{cg})
	\end{multlined}
}
{\arctan\frac{\omega_{cg}}{\omega_h}-\arctan
	\frac{\omega_{cg}}{\omega_b}}
\end{equation}
with $\nu^* \in [0,1]$ for CRONE-1 and

\begin{equation}\label{eq:nucrone2}
\nu^*=
\frac
{
	\begin{multlined}
	-\pi+M_\Phi+n_F\arctan\frac{\omega_{cg}}{\omega_F}+n_I(\frac{\pi}{2}-\arctan\frac{\omega_{cg}}{\omega_I})\\-\Phi_r(\omega_{cg})
	\end{multlined}
}
{\arctan\frac{\omega_{cg}}{\omega_h}-\arctan
	\frac{\omega_{cg}}{\omega_b}}
\end{equation}
with $\nu^* \in [1,2]$ for CRONE-2.

\subsection{Stability analysis}
Theorems \ref{th:1} \& \ref{th:2} can be used to guarantee asymptotic stability using the following $A_{cl}$-matrix:

\begin{equation}
A_{cl}=\begin{bmatrix}
\bar{A}&\bar{B}C_{nrp}\\
-B_{nrp}\bar{C}&A_{nrp}
\end{bmatrix}
\end{equation}
where $(\bar{A},\bar{B},\bar{C},\bar{D})$ are the state-space matrices of $\Sigma^*_r$ and $(A_{nrp}, B_{nrp},C_{nrp}, D_{nrp})$ are the state-space matrices of non-reset controller $\Sigma_{nr}$ and plant $\Sigma_p$ combined in series. $\bar{A}$ is defined as:
\begin{equation}
\bar{A}=\begin{bmatrix}
A_r&0\\0&A_r
\end{bmatrix}
\end{equation}
$\bar{B}$ and $\bar{C}$ are defined respectively as:
\begin{equation}
\bar{B}=\begin{bmatrix}
B_r\\B_r
\end{bmatrix},
\bar{C} = \begin{bmatrix}
pC_r&(1-p)C_r
\end{bmatrix}
\end{equation}
 and $A^*_\rho$ is defined as:
\begin{equation}
A^*_\rho=\mathrm{diag}(\bar{A_\rho},I_{n_r},I_{n_{nrp}})
\end{equation}
where $n_{nrp}$ is sum of the number of states of non-reset controller $n_{nr}$ and plant $n_p$. 

\section{Practical application}\label{sec:practical}

\color{black}
The idea of CRONE reset control is to obtain a robust controller capable of overcoming the robustness-performance trade-off. This is achieved by breaking Bode's gain-phase relation through introduction of nonlinearity. However, since the controller is analysed and designed using the pseudo-linear describing function approximation, the proposed controllers are tested on a precision positioning stage for validation.
\color{black}

\subsection{System overview}
\begin{figure}
	\centering
	\includegraphics[width=\linewidth]{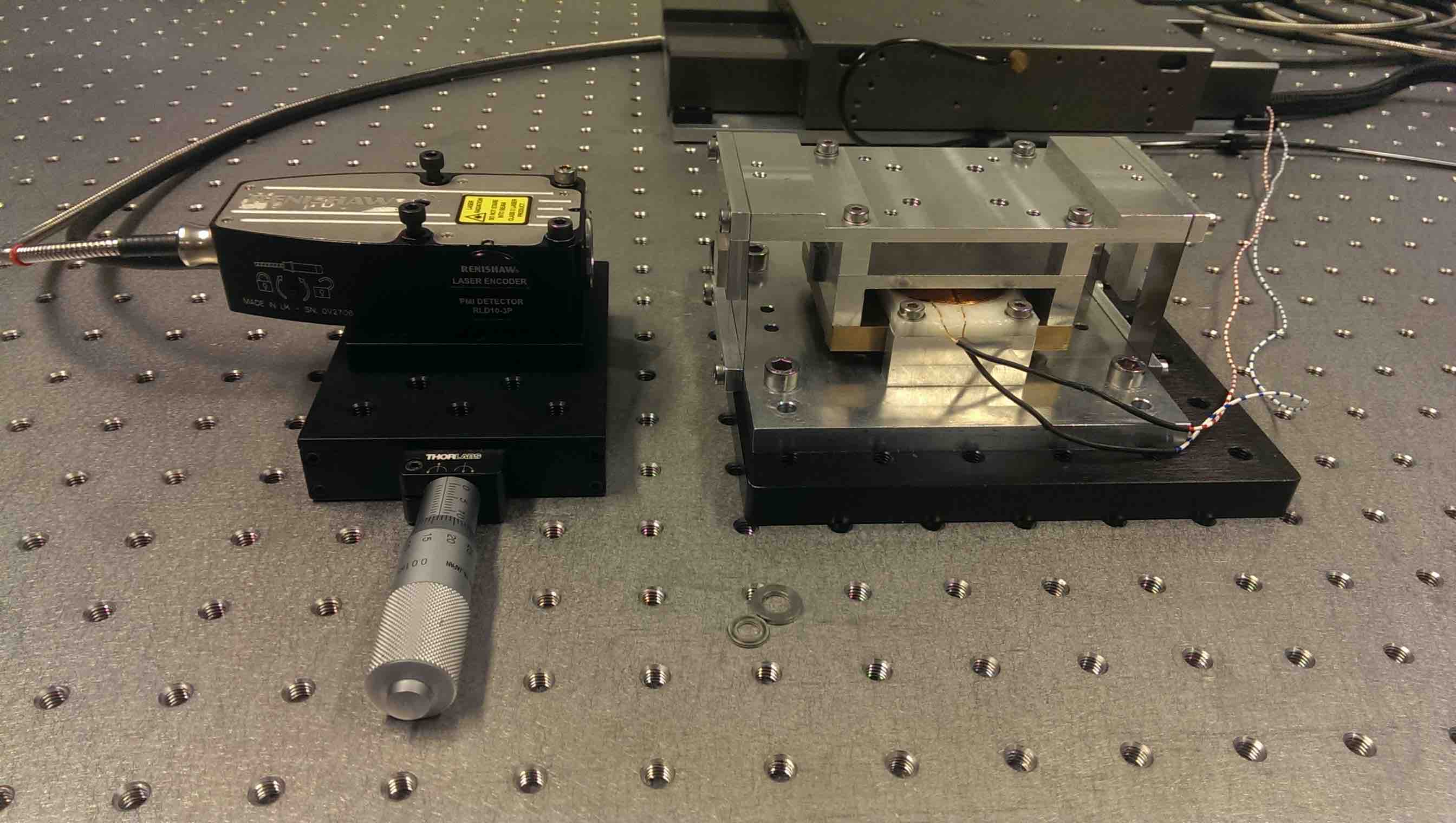}
	\caption{Picture of the Lorentz stage (right) with the laser encoder at the left.}
	\label{fig:setup}
\end{figure}

The system considered is a custom-designed one-degree-of-freedom nanometre precision positioning stage, actuated by a Lorentz actuator. This stage is linear-guided using flexures to attach the Lorentz actuator to the base of the stage and actuated at the centre of the flexures. With a \textit{Renishaw RLE10} laser encoder the position of the stage is read out with 10\si{\nano\metre} resolution. The setup is depicted in Fig. \ref{fig:setup}. All CRONE reset controllers are designed within a MATLAB/Simulink environment and implemented digitally via dSPACE DS1103 real-time control software with a sampling rate of 20\si{\kilo\hertz}.
The transfer function of this system is identified as:

\begin{equation}\label{eq:plant}
P(s)=\frac{0.5474}{0.5718s^2+0.95+146.3}e^{\num{-2.5e-4}s} .
\end{equation}

\begin{figure}
	\centering
	\includegraphics[width=\linewidth]{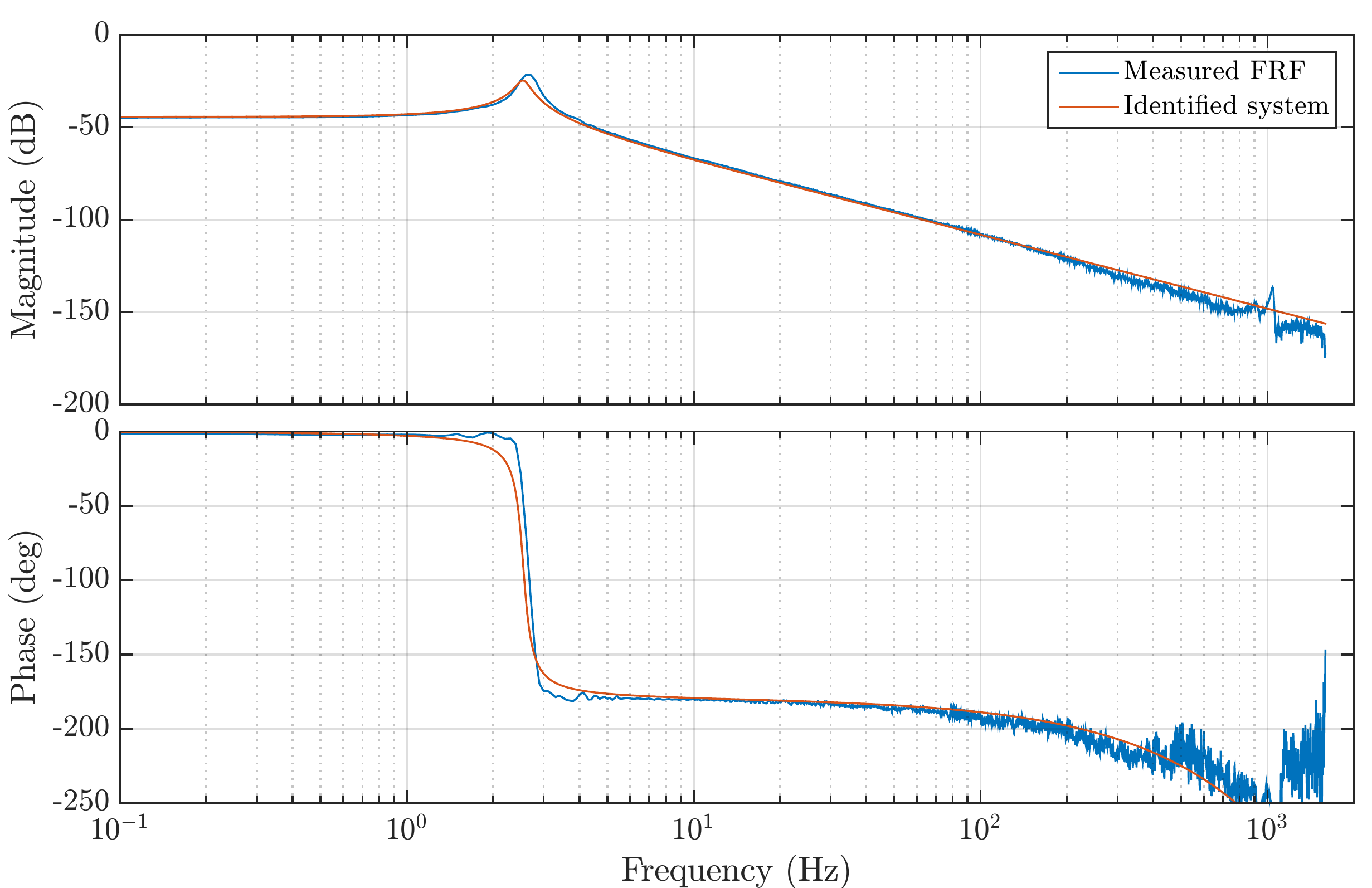}
	\caption{Frequency response of the system and the identified system 
		model.}
	\label{fig:frf}
\end{figure}

The frequency response of the system is shown in Fig. \ref{fig:frf} and shows the behaviour of a second order mass-spring-damper system with additional dynamics at higher frequencies and delay. 

%\begin{figure*}[tb]
%\centering
%\begin{subfigure}{0.45\linewidth}
%\centering
%\includegraphics[width=\linewidth]{setup3.jpg}
%\caption{}
%\label{fig:setup}
%\end{subfigure}
%\begin{subfigure}{0.45\linewidth}
%\centering
%\includegraphics[width=\linewidth]{fsfrf22}
%\caption{}
%\label{fig:frf}
%\end{subfigure}
%\caption{(\subref{fig:setup}) Picture of the Lorentz stage (right) with the laser encoder at the left. (\subref{fig:frf}) Frequency response of the system and the identified system 
%model.}
%\end{figure*}

\begin{table*}[htbp]
	\centering
	\caption{Parameters of CRONE-1 lag reset and CRONE-2 lag reset controllers.}
	\begin{tabular}{llll}\toprule
		Symbol&Parameter&Value CRONE-1& Value CRONE-2\\\midrule
		PM&phase margin&\SI{55}{\degree}&\SI{55}{\degree}\\
		$\omega_{cg}$&bandwidth&\SI{100}{\hertz}&\SI{100}{\hertz}\\
		$\omega_b$&lead corner frequency&\SI{12.5}{\hertz}&\SI{12.5}{\hertz}\\
		$\omega_h$&lag corner frequency&\SI{800}{\hertz}&\SI{800}{\hertz}\\
		$\omega_I$&integrator corner frequency&\SI{8.33}{\hertz}&\SI{8.33}{\hertz}\\
		$\omega_F$&low-pass filter corner frequency&\SI{1200}{\hertz}&\SI{1200}{\hertz}\\
		$n_I$&integrator order&1&2\\
		$n_F$&low-pass filter order&1&3\\
		$N$&Oustaloup approximation order&4&4\\
		$p$&reset percentage&0.5&0.5\\
		$\gamma$&partial reset&0.5&0.5\\\bottomrule
	\end{tabular}
	\label{tab:parameterscrone}
\end{table*}

\begin{figure*}[htbp]
	\centering
	\begin{subfigure}{0.48\linewidth}
		\includegraphics[width=\linewidth]{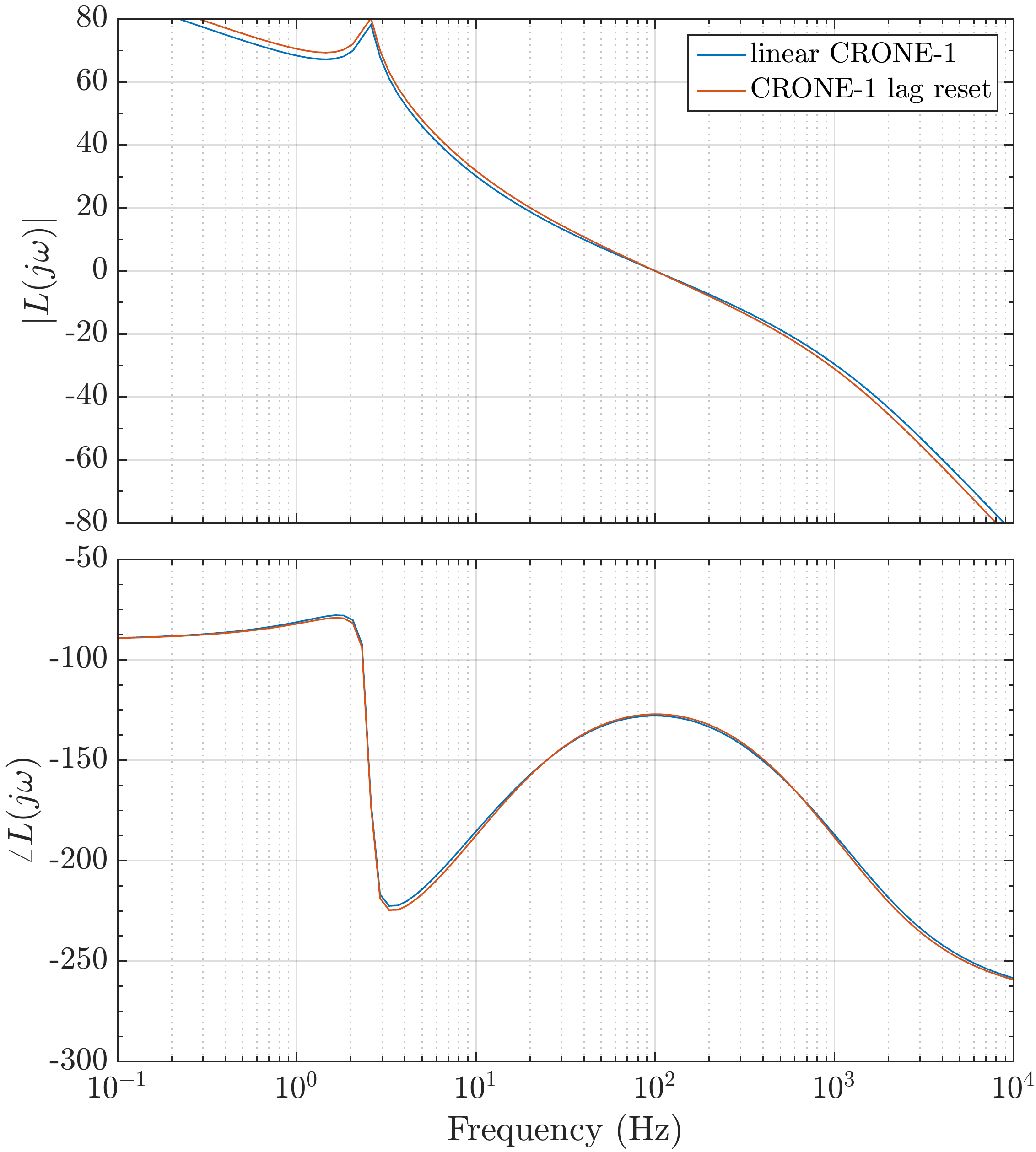}
		\caption{}
		\label{fig:crone1llol}
	\end{subfigure}
	\begin{subfigure}{0.48\linewidth}
		\includegraphics[width=\linewidth]{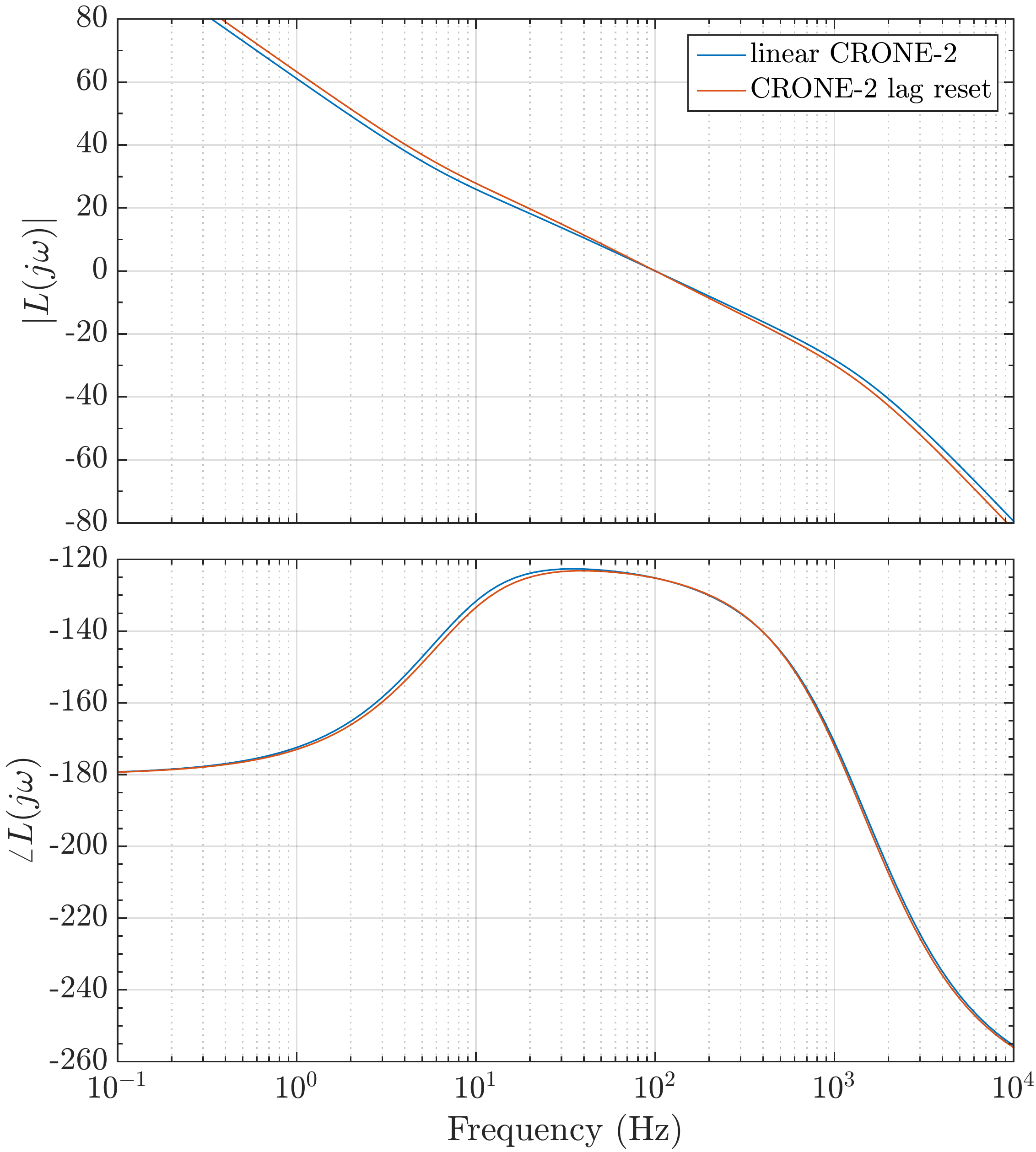}
		\caption{}
		\label{fig:crone2llol}
	\end{subfigure}
	\caption{Theoretical open-loop responses linear CRONE versus CRONE reset for (\subref{fig:crone1llol}) CRONE-1 lag reset and (\subref{fig:crone2llol}) CRONE-2 lag reset. The expected responses of CRONE reset are shown for $\gamma = p = 0.5$ as provided in Table. \ref{tab:parameterscrone}}
	\label{fig:ol}
\end{figure*} 

\subsection{Controller design}

\color{black}
The design of CRONE reset controllers for testing and validation provides us with choices over part of controller being reset as well as over the values of $p$ and $\gamma$, resulting in infinite possible choices. In this paper, we have chosen resetting lag part and also $p = \gamma = 0.5$ for validation. Both CRONE-1 and CRONE-2 reset controllers are designed for the above-described positioning stage. The complete set of parameters of the two controllers can be found in Tab. \ref{tab:parameterscrone}. Linear CRONE controllers are also designed with either $p = 1$ or $\gamma = 1$ or both, since all combinations result in linear CRONE design.
\color{black}

The theoretical open-loop responses are plot against its linear base equivalents in Fig. \ref{fig:ol}. Within these graphs, it is evident that for the same phase margin, the open-loop frequency response of CRONE reset provides a better open loop shape i.e. higher gain at low frequency which should result in better tracking. 

%As a confirmation of equal phase margins, overshoot of simulated step responses are analysed and shown in Fig. \ref{fig:isodamping}, as overshoot is a direct measure of phase margin. The simulations are done for controllers sampled with a frequency of \SI{20}{\kilo\hertz}. It can be seen that in the case of CRONE-1 lag reset, overshoot and thus a similar phase margin of \SI{55}{\degree} is maintained. In the case of CRONE-2 lag reset, the  overshoot corresponds with the overshoot of a linear CRONE-2 controller with phase margin of \SI{52.4}{\degree}. Since reset phase lead in the latter case is \SI{10.5}{\degree} using (\ref{eq:phirllag}), still some additional phase is expected to be attained using reset in the CRONE-2 lag reset controlled system.
%
%\begin{figure}[htb]
%\centering
%\begin{subfigure}{0.45\linewidth}
%\includegraphics[width=\linewidth]{stepcrone1}
%\caption{}
%\label{fig:stepcrone1}
%\end{subfigure}
%\begin{subfigure}{0.45\linewidth}
%\includegraphics[width=\linewidth]{stepcrone2}
%\caption{}
%\label{fig:stepcrone2}
%\end{subfigure}
%\caption{Simulated step responses }
%\label{fig:isodamping}
%\end{figure}

\section{Experimental results and discussion}\label{sec:results}
Experimental data has been retrieved for two purposes, namely: showing improvement in frequency domain through sensitivity functions and thus applicability of describing function analysis, and; showing improvement in time domain proving better robustness-performance trade-offs of CRONE reset control compared to linear CRONE control. 

\subsection{Frequency domain results}\label{sec:waterbed}

\begin{figure}[!htb]
	\centering
	\includegraphics[width=\linewidth]{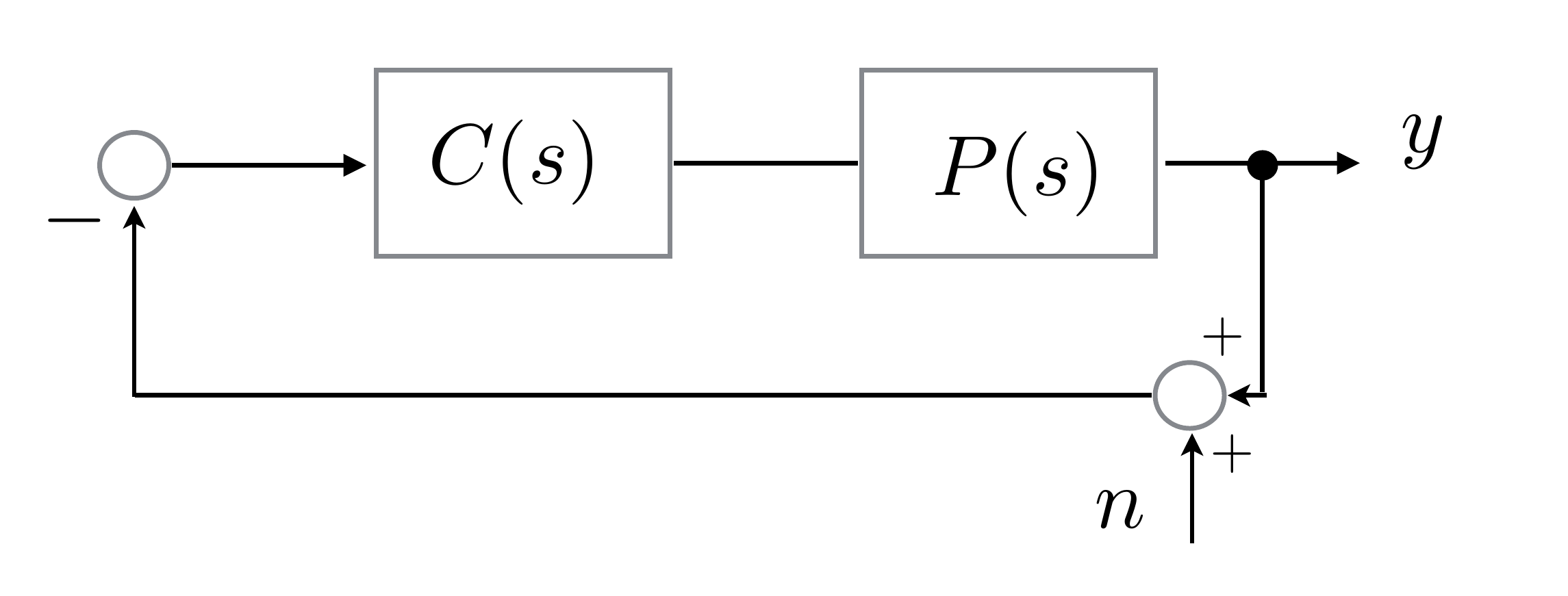}
	\caption{Block diagram of the control loop and signals used for identification of $T(j\omega)$ and $S(j\omega)$.}
	\label{fig:blockdiagsens}
\end{figure}

Complementary sensitivity function  $T(j\omega)$ and sensitivity function $S(j\omega)$ were identified using a frequency sweep at the position of signal $n$ in {Fig. \ref{fig:blockdiagsens}}. $T(j\omega)$ then can be identified as the transfer from $-n$ to $y$, whereas $S(j\omega)$ is identified as the transfer from $n$ to $y+n$. The frequency sweep was done from \SI{0.1}{\hertz} to \SI{2.5}{\kilo\hertz} with a target time of \SI{120}{\second} and a total duration of \SI{480}{\second}.

\begin{figure*}[htbp]
	\centering
	%\begin{subfigure}{0.45\linewidth}
	%\includegraphics[width=\linewidth]{cronestep1}
	%\caption{}
	%\label{fig:cronestep1}
	%\end{subfigure}
	\begin{subfigure}{0.48\linewidth}
		\includegraphics[width=\linewidth]{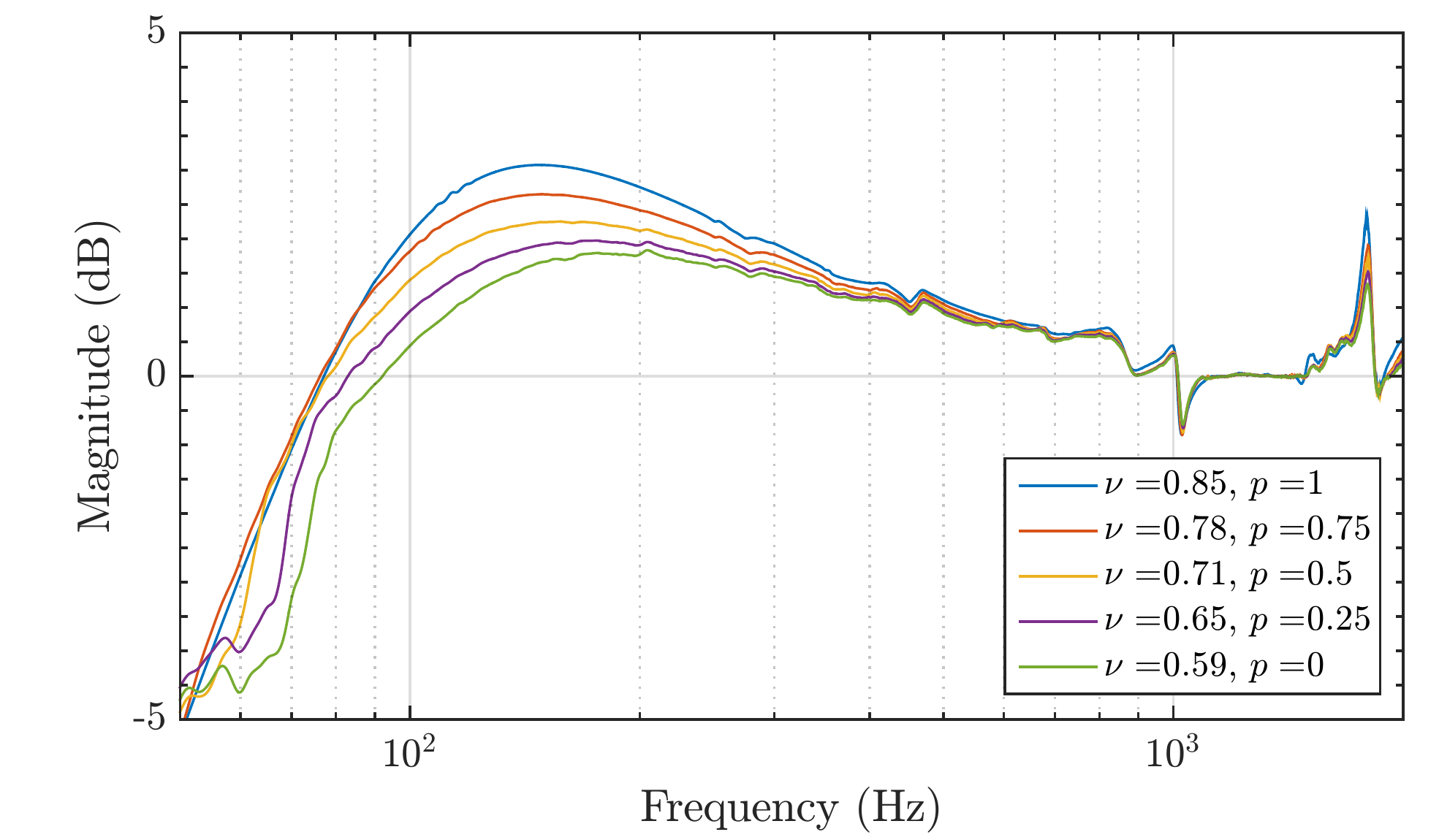}
		\caption{}
		\label{fig:crone1swater}
	\end{subfigure}
	\begin{subfigure}{0.48\linewidth}
		\includegraphics[width=\linewidth]{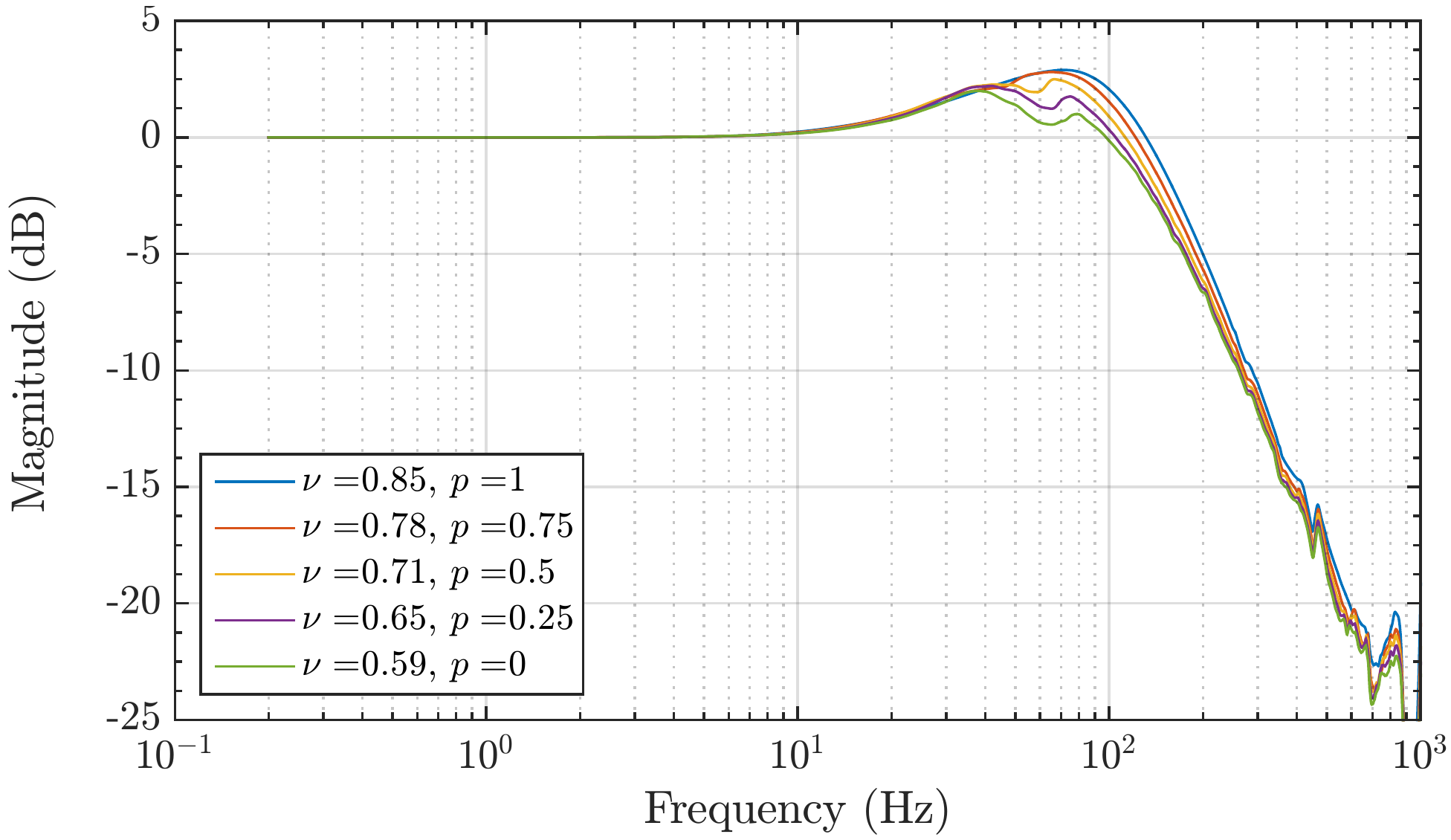}
		\caption{}
		\label{fig:crone1cswater}
	\end{subfigure}
	\begin{subfigure}{0.48\linewidth}
		\includegraphics[width=\linewidth]{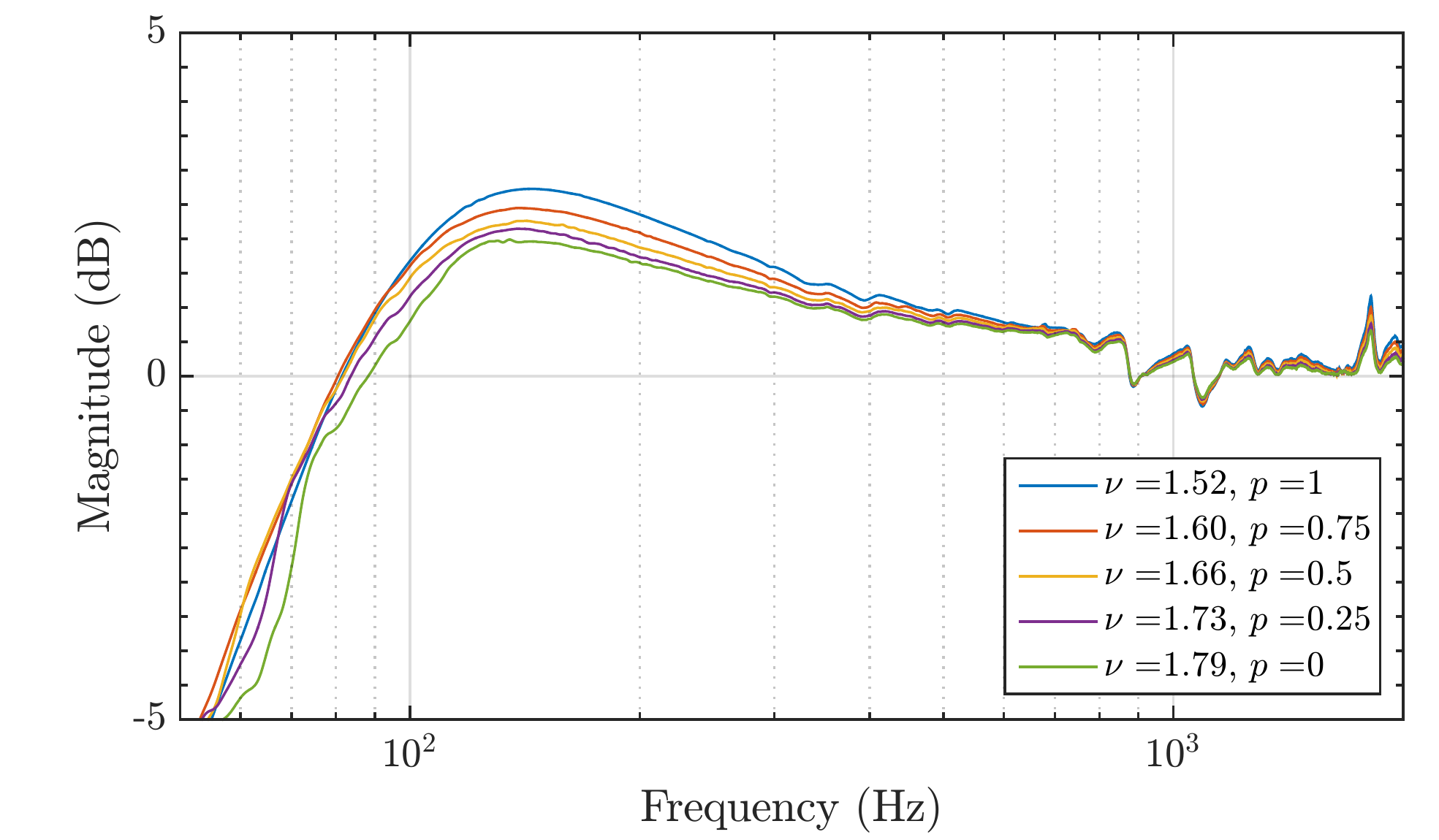}
		\caption{}
		\label{fig:crone2swater}
	\end{subfigure}
	\begin{subfigure}{0.48\linewidth}
		\includegraphics[width=\linewidth]{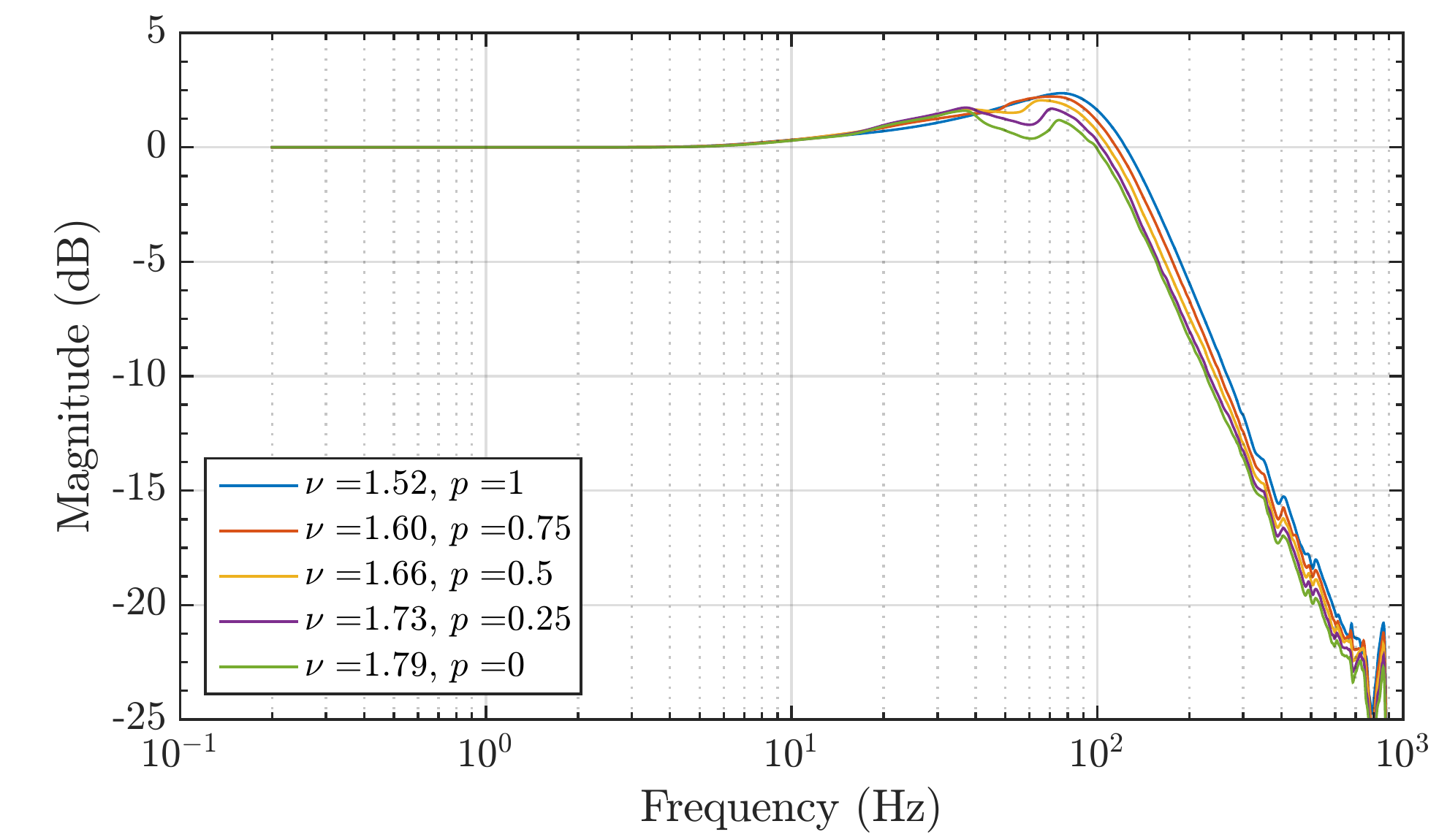}
		\caption{}
		\label{fig:crone2cswater}
	\end{subfigure}
	\caption{Measured frequency responses for CRONE-1 lag reset (\subref{fig:crone1swater}) sensitivity function, (\subref{fig:crone1cswater}) complementary sensitivity function and CRONE-2 lag reset (\subref{fig:crone2swater}) sensitivity function and (\subref{fig:crone2cswater}) complementary sensitivity function for $\gamma=0.5$ and different $p$-values.}
	\label{fig:sensfuncwater}
\end{figure*}

The identified frequency responses $S(j\omega)$ and $T(j\omega)$ for CRONE-1 lag reset and CRONE-2 lag reset for $\gamma=0.5$ and different values of $p$ are shown in Fig. \ref{fig:sensfuncwater}. In the complementary sensitivity functions in Fig. \ref{fig:crone1cswater} and Fig. \ref{fig:crone2cswater} it can be seen that the peak value reduces for decreasing value of $p$ (increasing non-linearity in system).  Additionally there is better attenuation of high frequencies. Both contribute to attaining better reference tracking performance. 

\color{black}
In Fig. \ref{fig:crone1swater} and Fig. \ref{fig:crone2swater} the $S(j\omega)$ is shown for the same controller parameters. Here it is evident that there is a gain reduction of both the peak value and sensitivity at higher frequencies. The low frequency sensitivity functions are not shown for reasons of inevitable low coherence. The results presented above indicate an improvement in the robustness-performance trade-off as decrease of both gain peak values and gain at low frequencies is observed. 
\color{black}

\subsection{Time domain results: reference-tracking}
\begin{figure}[htbp]
\centering
\includegraphics[width=\linewidth]{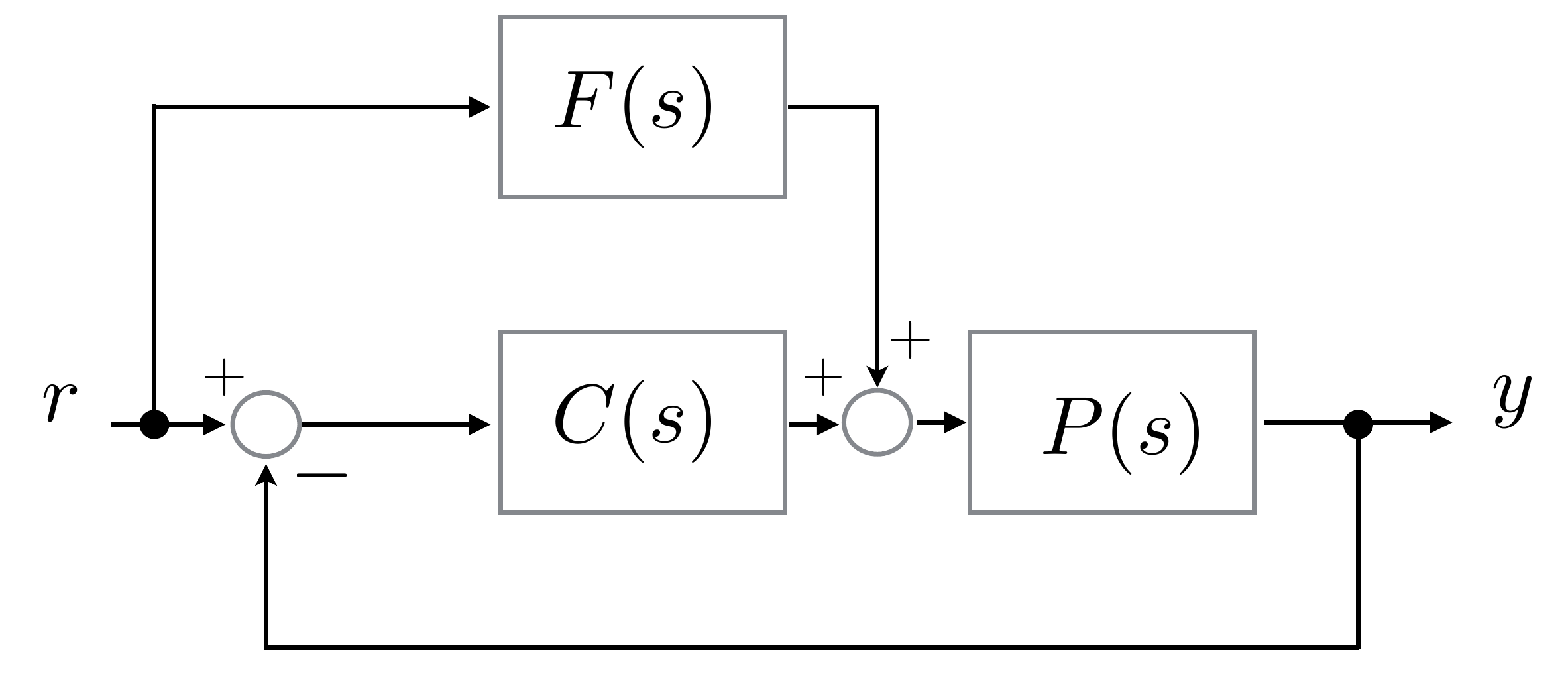}
\caption{Block diagram of the controlled system.}
\label{fig:blockdiag}
\end{figure}

Fourth order trajectory planning as in \cite{lambrechts2005} is used to compute a triangular wave reference signal. This reference signal is representative of scanning motions in precision wafer stages. Additionally, second order feedforward as formulated by the same authors has been implemented. The feedforward controller provides a feedforward force $F$ that is computed as:

\begin{equation}
F=ma+cv
\end{equation}
in which $m$ is the stage mass, $c$ is the damping coefficient and $a$ and $v$ are the acceleration and velocity of the stage respectively.  In Fig. \ref{fig:blockdiag} a common feedforward controller $F(s)$ is used with linear CRONE and CRONE reset controllers as feedback controllers $C(s)$ for performance comparison.

\begin{figure*}[htbp]
\centering
\begin{subfigure}{0.45\linewidth}
\includegraphics[width=\linewidth]{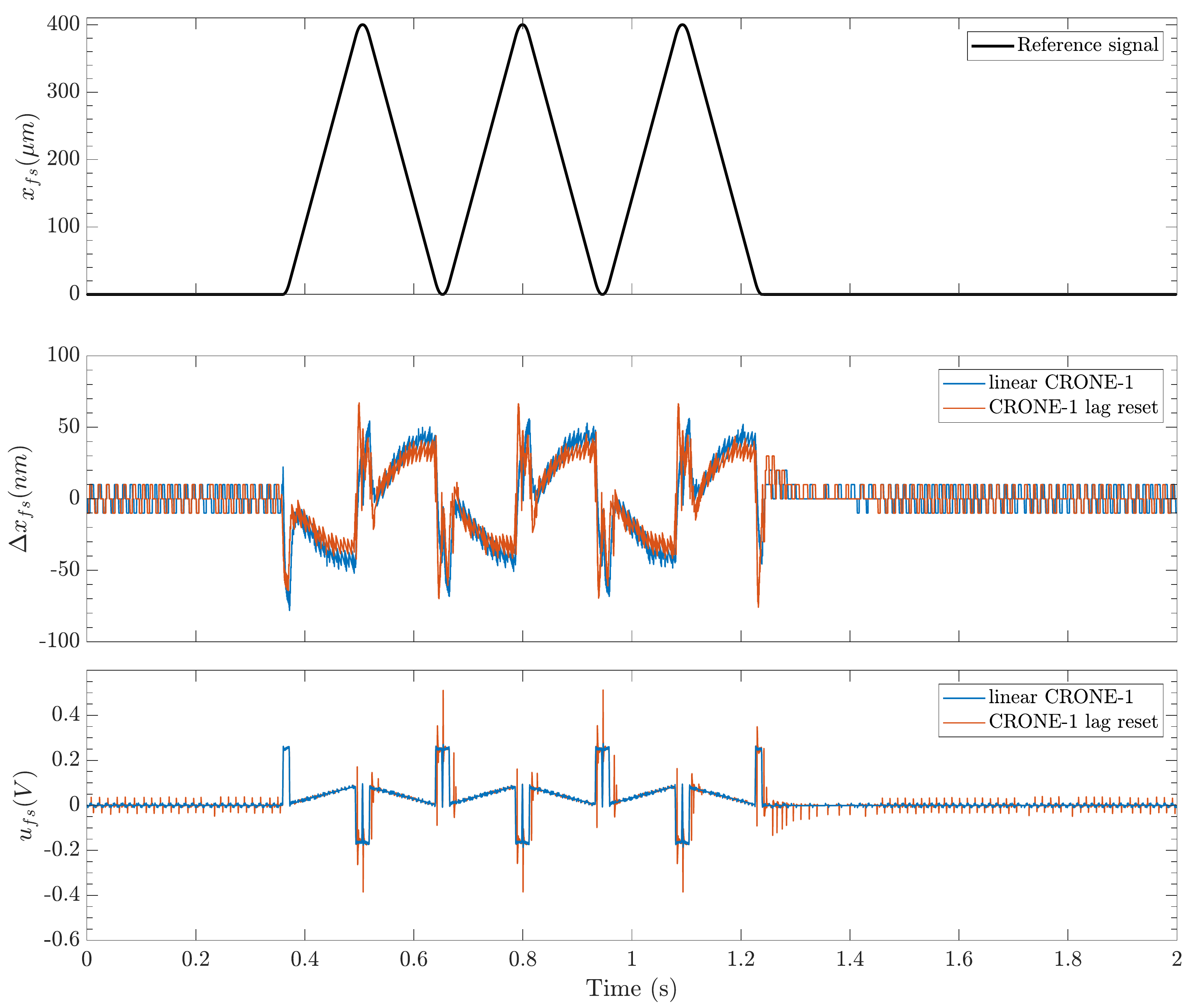}
\caption{}
\label{fig:triangcrone1}
\end{subfigure}
\begin{subfigure}{0.45\linewidth}
\includegraphics[width=\linewidth]{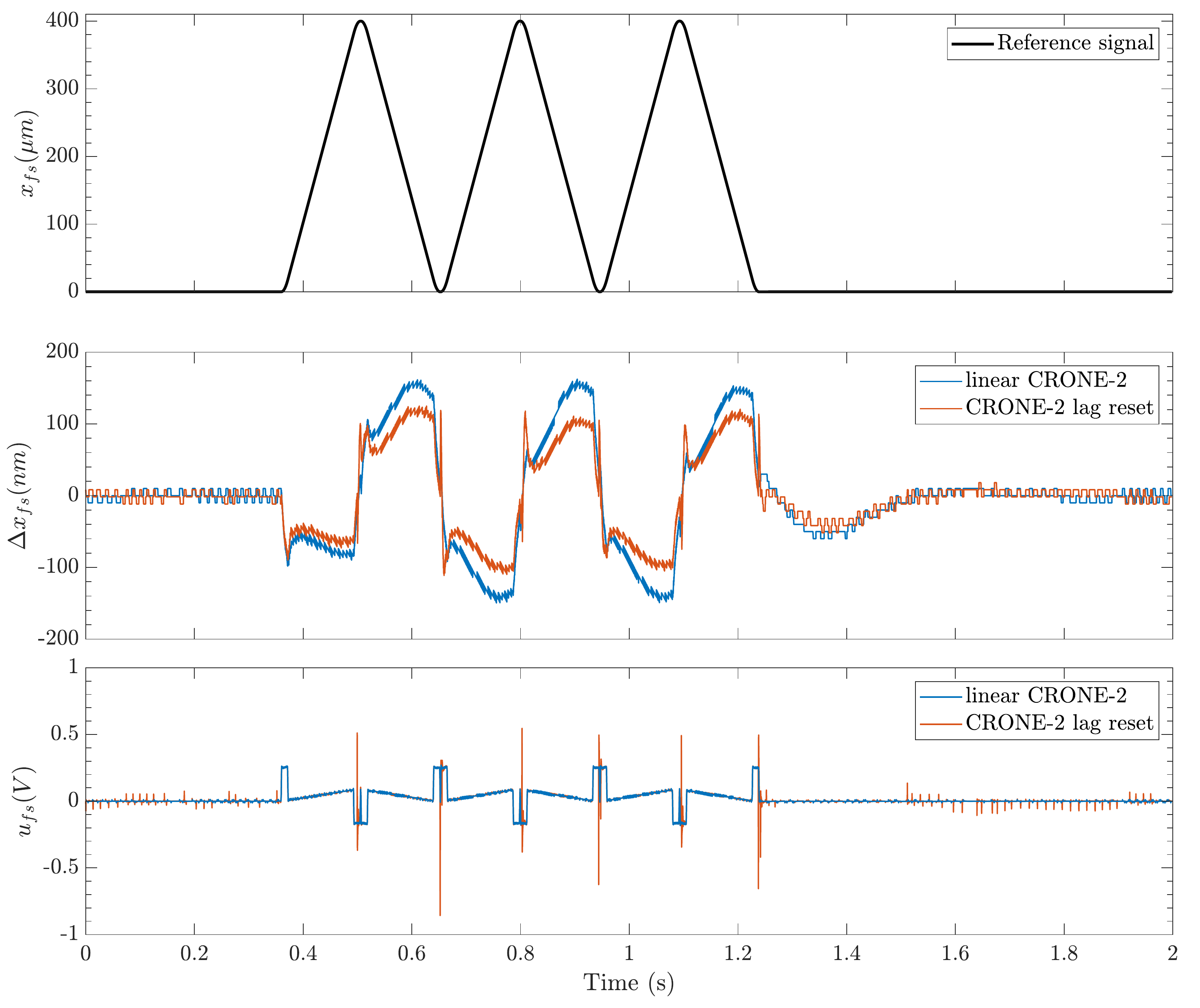}
\caption{}
\label{fig:triangcrone2}
\end{subfigure}
\caption{Reference-tracking of a fourth order input-shaped triangular wave signal for (\subref{fig:triangcrone1}) CRONE-1 lag reset and (\subref{fig:triangcrone2}) CRONE-2 lag reset compared to linear CRONE for the same phase margin.}
\label{fig:triang}
\end{figure*}

The tracking errors for the CRONE lag reset controllers are compared to linear CRONE controllers for the same phase margin in Fig. \ref{fig:triang}. According to theory, as has been explained in section \ref{sec:cronereset}, the CRONE reset system will perform better than the linear CRONE reset system with similar phase margin. This is confirmed by the measurements and is evident from the error plots of Fig. \ref{fig:triang}. For CRONE-1 lag reset, the RMS error reduces from \SI{22.1}{\nano\metre} to \SI{19.6}{\nano\metre} compared to the linear CRONE-1 controller. For CRONE-2 lag reset, RMS error is reduced from \SI{70.7}{\nano\metre} to \SI{52.8}{\nano\metre}. Hence, better performance in terms of reference-tracking has been achieved in CRONE reset control compared to linear CRONE. This improvement in tracking performance is also in line with the $S(j\omega)$ graphs in Fig. \ref{fig:crone1swater} and Fig. \ref{fig:crone2swater} which is basically an estimate of the error wrt reference.

It has to be noted that maximum control effort peaks are larger in the non-linear system. Nevertheless, if the system does not suffer from control saturation, this increase is no problem.

\color{black}

\subsection{Time domain results: noise-attenuation}
The noise-attenuation performance of the CRONE lag reset and linear CRONE reset controllers is evaluated using the system response to a sine noise input signal. Noise power is calculated for sine signal of different frequencies above the bandwidth for a duration of 5s. The decrease in average power in decibels for $p=\gamma=0.5$ with respect to linear case is shown in Tables \ref{tab:noise} and \ref{tab:noise2}. In literature, one of the most noted advantages of reset is its improved performance in noise attenuation. This property of reset is validated again in the case of CRONE reset with these results. These results provide further validation of $T(j\omega)$ plotted in Fig. \ref{fig:crone1cswater} and Fig. \ref{fig:crone2cswater} which is an estimate of error due to noise at a range of frequencies.

\begin{table}[!htb]
	\color{black}
\caption{\color{black}Reduction of average power of noise response for $\gamma=0.5$, $p=0.5$ with respect to linear case for CRONE-1}
\centering
\begin{tabular}{ll}\toprule
Frequency (\si{\hertz})&noise reduction (\si{\decibel})\\\midrule
300&2.46\\
400&2.74\\
500&2.72\\
600&2.98\\
700&2.66\\
800&2.57\\
900&1.79\\
1000&2.59\\\bottomrule
\end{tabular}
\label{tab:noise} 
\end{table}

\begin{table}[!htb]
	\color{black}
	\caption{\color{black}Reduction of average power of noise response for $\gamma=0.5$, $p=0.5$ with respect to linear case for CRONE-2}
	\centering
	\begin{tabular}{ll}\toprule
		Frequency (\si{\hertz})&noise reduction (\si{\decibel})\\\midrule
		300&2.94\\
		400&3.14\\
		500&3.55\\
		600&3.14\\
		700&3.30\\
		800&3.10\\
		900&3.93\\
		1000&2.92\\\bottomrule
	\end{tabular}
	\label{tab:noise2} 
\end{table}

\color{black}

%\begin{figure}[!htb]
%\centering
%\includegraphics[width=\linewidth]{noise1k2}
%\caption{Measured response for sine noise signal with frequency of 1\si{\kilo\hertz} and amplitude of 2\si{\micro\metre}.}
%\label{fig:noise1k}
%\end{figure}

\color{black}
\subsection{Time domain results: disturbance rejection}

\begin{figure*}[htbp]
	\centering
	\begin{subfigure}{0.45\linewidth}
		\includegraphics[width=\linewidth]{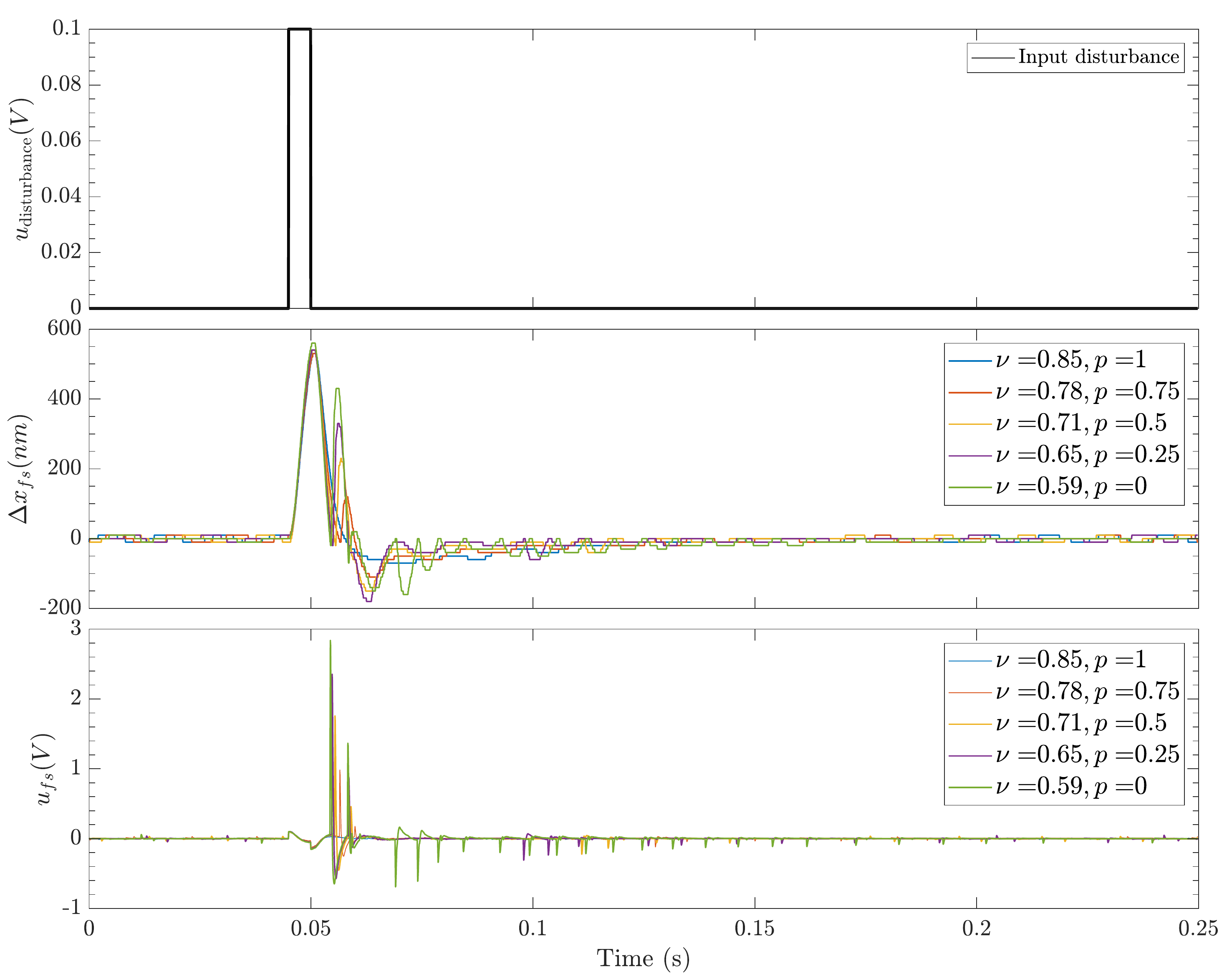}
		\caption{}
		\label{fig:distcrone1}
	\end{subfigure}
	\begin{subfigure}{0.45\linewidth}
		\includegraphics[width=\linewidth]{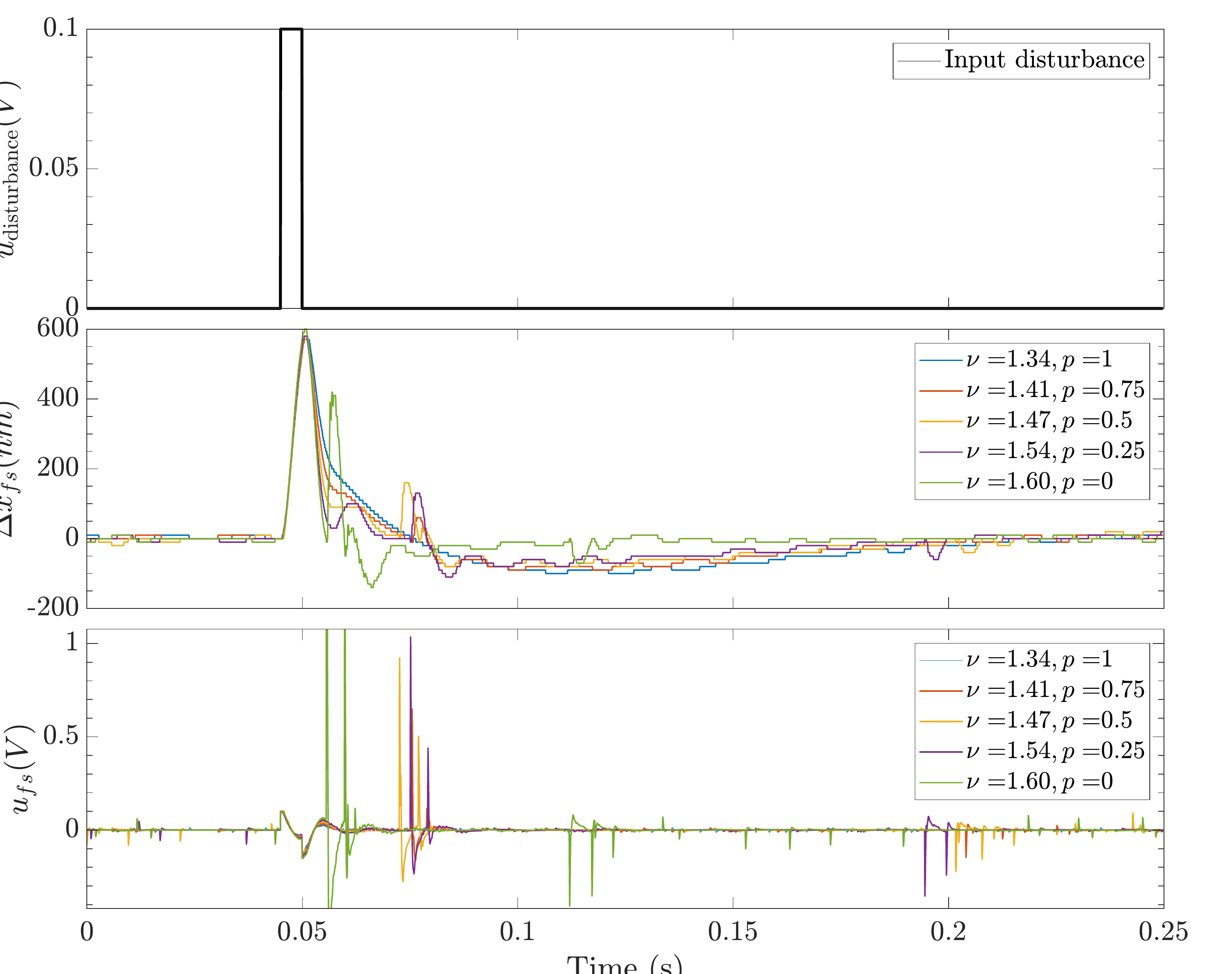}
		\caption{}
		\label{fig:distcrone2}
	\end{subfigure}
	\caption{Response to a pulse disturbance for lag reset with $\gamma = 0.5$ for 
		(\subref{fig:distcrone1}) CRONE-1 lag reset and (\subref{fig:distcrone2}) CRONE-2 lag reset.}
	\label{fig:distrej}
\end{figure*}

A large percentage of the high precision achieved in the high-tech industry is due to the accurate design of the feedforward controller. In several cases, the feedforward controller and the reference itself are designed so well that the feedback controller does not see any error due to change in reference. In such a scenario, the feedback controller is mainly responsible for noise attenuation and disturbance rejection. The improved noise attenuation performance of CRONE reset controllers has already been validated in the previous subsection. The response of CRONE reset controllers for a step disturbance are shown in Fig. \ref{fig:distrej}. While the maximum error due to this disturbance is not significantly different in the responses of CRONE reset and linear CRONE, significant difference is seen in the settling time. The settling time is defined and calculated as the time required to decrease to $15\%$ of the maximum peak value. The computed settling times are shown in Tables \ref{tab:dist} and \ref{tab:dist2} for CRONE-1 reset and CRONE-2 reset respectively.

%\begin{figure*}[htbp]
%	\centering
%	\includegraphics[width=\linewidth]{Distrej}
%	\caption{Response to a step disturbance for lag-lead reset with g = 0.5 for (a)
%		CRONE-1 and (b) CRONE-2.}
%	\label{fig:distrej}
%\end{figure*}

\begin{table}[!htb]
	\color{black}
	\caption{\color{black}Settling time (Time to reach $15\%$ of peak value $\hat{x}_{fs}$) for CRONE-1 reset controllers with value of $\gamma$ fixed to 0.5}
	\centering
	\begin{tabular}{lll}\toprule
		p & $\hat{x}_{fs}$ ()\SI{}{\nano\metre}) & settling time (ms)\\
		1&530&55.8\\
		0.75&540&65.5\\
		0.5&540&65.5\\
		0.25&550&66.0\\
		0&560&76.8\\\bottomrule
	\end{tabular}
	\label{tab:dist} 
\end{table}

\begin{table}[!htb]
	\color{black}
	\caption{\color{black}Settling time (Time to reach $15\%$ of peak value $\hat{x}_{fs}$) for CRONE-2 reset controllers with value of $\gamma$ fixed to 0.5}
	\centering
	\begin{tabular}{lll}\toprule
		p & $\hat{x}_{fs}$ ()\SI{}{\nano\metre}) & settling time (ms)\\
		1&570&146.0\\
		0.75&570&135.7\\
		0.5&570&128.1\\
		0.25&590&125.1\\
		0&600&68.3\\\bottomrule
	\end{tabular}
	\label{tab:dist2} 
\end{table}

In the case of disturbance rejection, it is seen that the performance of the controllers does not match the expectation from describing function. While in the case of CRONE-2 reset controllers, a reduction in settling time is seen, there is a slight increase in max displacement due to disturbance. However, in the case of CRONE-1 significant rise in settling time is also seen. This could possibly be explained as effect of higher order harmonics introduced by reset. These results also show that while describing functions are reliable in estimating performance improvement in some cases, they are not reliable under all circumstances. This discrepancy is even more evident when we consider that process sensitivity which is the estimate of error wrt disturbance can be obtained in linear systems by multiplying sensitivity with the plant being controlled. The sensitivity $S(j\gamma)$ for the designed controllers is plotted in Fig. \ref{fig:crone1cswater} and Fig. \ref{fig:crone2cswater}. Since the plant is constant for all the designed controllers, the disturbance rejection performance should follow the describing function estimation. However, this is not the case as seen from the practical results. Hence, new frequency domain tools capable of accurately estimating closed-loop performance of reset systems are required. 

\color{black}

\section{Conclusion}\label{sec:conclusion}
The first part of this work has a theoretical focus: novel and general design rules are developed in the synthesis of the proposed CRONE reset controller. These general rules are applicable to a broad range of reset strategies that can be taken for both first generation CRONE reset and second generation CRONE reset. The developed theory was used in the design of a CRONE-1 lag reset and CRONE-2 lag reset controller. For these controllers it was shown that for similar phase margin, better open-loop shape can be achieved compared to linear CRONE control, thus providing relief from Bode's fundamental gain-phase relation and fundamental robustness-performance trade-offs. 

In the second part of this paper, the designed CRONE reset controllers have been validated on a Lorentz-actuated precision stage. Firstly, it was shown that the sensitivity function and complementary sensitivity function, which were identified from measurement data, improve in the frequency range of interest. Both sensitivity and complementary sensitivity peaks reduced as well as gain at high frequencies. Using time domain reference-tracking results for a fourth order input-shaped triangular reference signal, it was shown that the better open-loop shape of CRONE-reset indeed improves reference-tracking performance. For both CRONE-1 lag reset and CRONE-2 lag reset, reduction of RMS tracking error was observed. The use of reset for improved noise rejection performance is also validated on the practical setup.

\color{black}

The reliability of describing function however is questionable in the case of disturbance rejection. While in the case of CRONE-2 reset improvement is seen in settling time, performance deteriorates both in terms of maximum displacement and settling time for CRONE-1 reset.

Several challenges of reset control are not addressed in this paper but will be considered in future work. These include limit cycles, low-frequency disturbances and higher order harmonic behaviour amongst others. As seen with the results of disturbance rejection, existing analysis of open-loop and closed-loop behaviour of reset systems using describing function is insufficient under certain conditions. This requires new tools and methodologies for frequency domain analysis of reset systems. However, the results shown in this paper  are already promising: for the designed CRONE reset controllers with partial reset and reset percentage already improve tracking performance and noise attenuation with respect to linear CRONE. This means that with future study into the effect of higher order harmonics, performance of CRONE reset can be further improved. Also such a study of higher order harmonics will provide more insight into the choice of values of $p$ and $\gamma$. While $0.5$ has been chosen as the value for comparison and validation in this paper, the best value to achieve required specifications for any system can be accurately chosen when the complete closed-loop performance including the effect of higher order harmonics can be predicted.

\color{black}

\bibliographystyle{IEEEtran}
\bibliography{references}

\end{document}